\documentclass[11pt, epsfig]{article}
\usepackage{epsfig, amsmath, amssymb, amsthm, times}
\usepackage{graphicx}
\usepackage{color}

\newcommand{\R}{{\mathbb R}}

\newcommand{\Z}{{\mathbb Z}}
\newcommand{\T}{{\mathbb T}}

\usepackage[mathcal]{eucal}

\theoremstyle{defintion}

\theoremstyle{remark}

\theoremstyle{example}

\theoremstyle{assumption}

\def\P{{\mathbb P~}}
\def\SS{{\mathbb S}}
\def\T{{\mathbb T}}

\def\lbl{\label}
\def\be{\begin{equation}}
\def\ee{\end{equation}}
\def\p{\partial}

\newcommand{\1}{{i\mkern1mu}}

\title{Stability of clusters in the second--order Kuramoto model on random graphs}
\author{Georgi S. Medvedev\thanks{Department of Mathematics, 
		Drexel University, 3141 Chestnut Street, Philadelphia, PA 19104,
		{\tt medvedev@drexel.edu}} \and Mathew S. Mizuhara\thanks{Department of 
			Mathematics and Statistics,
			The College of New Jersey,
			{\tt  mizuharm@tcnj.edu}}}
\begin{document}
\maketitle
\begin{abstract}
  The Kuramoto model of coupled phase oscillators with inertia on Erd{\H o}s-R{\' e}nyi
  graphs is analyzed in this work. For a system with intrinsic frequencies sampled from a
  bimodal distribution we identify a variety of two cluster patterns
  and study their stability. To this end, we decompose the description of the cluster dynamics
  into two systems: one governing the (macro) dynamics of
  the centers of mass of the two clusters and the second governing
  the (micro) dynamics of individual oscillators inside each cluster.
  The former is a 
  low-dimensional ODE  whereas the latter is a system of two coupled
  Vlasov PDEs. Stability of the cluster dynamics depends on the stability
  of the low-dimensional group motion and on coherence of the oscillators
  in each group. We show that the loss of coherence in one of the clusters leads
  to the loss of stability of a two-cluster state and to
  formation of chimera states. The analysis of this paper can be generalized
  to cover states with more than two clusters and to coupled systems on W-random graphs.
  Our results apply to a model of a power grid with fluctuating sources.
\end{abstract}
\section{Introduction}
\setcounter{equation}{0}

Understanding principles underlying collective behavior in large networks of interacting dynamical systems 
is an important problem with applications ranging from neuronal networks to power grids. Many dynamical
models on networks have been proposed to this effect. The Kuramoto model (KM) of coupled phase oscillators
has had a widespread success
due to its analytical simplicity and universality of the dynamical mechanisms that it helped to
reveal. It describes the evolution of interconnected phase oscillators $u_{n,i}:\R^+\to\R/2\pi\Z$
having intrinsic frequencies $\omega_{n,i}$:
\be\lbl{KM}
\dot u_{n,i}=\omega_{n,i} +Kn^{-1}\sum_{j=1}^n  a_{n,ij} \sin\left(u_{n,j}-u_{n,i}+\alpha\right),\quad i\in [n].
\ee
The sum on the right--hand side models the interactions between the oscillators, $\alpha\in [0,2\pi)$ determines
the type of interactions (attractive vs repulsive), and $K$ is the 
strength of coupling.
The spatial structure of interconnections is encoded in the adjacency matrix $(a_{n,ij})$.
The KM plays an important role in the theory of synchronization. We mention two major contributions that are
especially relevant to the present study. First, it reveals a universal mechanism for the transition to synchronization in
systems of coupled oscillators
with random intrinsic frequencies. The analysis of the KM shows that there is a critical value of the coupling
strength $K_c$ separating the incoherent (mixing) dynamics (Fig.~\ref{f.1}{\bf a}) from synchronization (Fig.~\ref{f.1}{\bf b})
\cite{StrMir91, Chi15a, ChiMed19a}. Second, studies of the KM led to the discovery of chimera states, patterns 
combining regions of coherent and incoherent dynamics \cite{KurBat02, AbrStr06, Ome18}.

Having reviewed the classical KM, we now turn to its generalization that is the
main focus of this paper:
\be\lbl{2KMa}
\ddot u_{n,i} +\gamma \dot u_{n,i}=\omega_{n,i} +K n^{-1}
\sum_{j=1}^n a_{n,ij} \sin\left(u_{n,j}-u_{n,i}+\alpha\right),\quad i\in [n].
\ee
The main new additions here are the second-order terms. 
The other parameters are
the damping constant $\gamma >0$ and the random torques $\omega_{n,i}$, which 
we keep referring to as intrinsic frequencies to emphasize the parallels with the 
classical KM \eqref{KM}. The system of equations \eqref{2KMa} can be viewed a model of coupled
pendula. Systems of equations like  \eqref{2KMa} are widely 
used for modeling power networks \cite{DB12, SMV84}.
The inclusion of the second order terms  makes the dynamics substantially more 
complex \cite{AntRuf, TSL97, JBL18}. In particular, the second order model is known for its capacity to
generate a rich variety of coherent clusters \cite{TSL97,BBB16}. Clusters exist for different types
of connectivity and different probability distributions of intrinsic frequencies.
We experimented with uniform, Gaussian, and certain multimodal distributions
and used all-to-all and random Erd{\H o}s-R{\' e}nyi (ER) connectivity. In each case, we saw an abundance
of clusters (Fig.~\ref{f.2}). Furthermore, we often found multiple cluster states
coexisting for the same values of parameters (Fig.~\ref{f.3}).
\begin{figure}
\begin{center}
\textbf{a})\includegraphics[height = .3\textwidth, width = .3\textwidth]{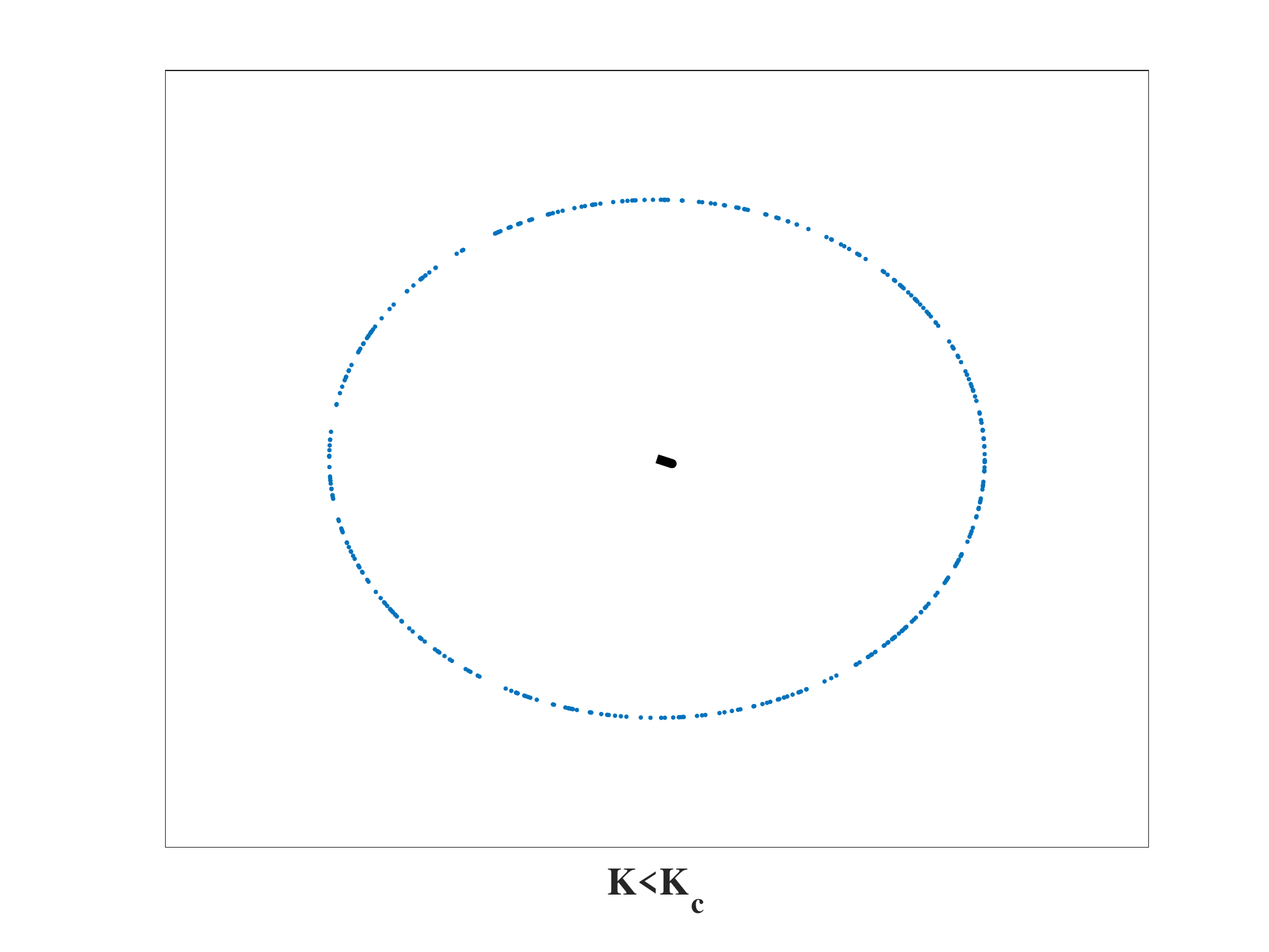}\;
\textbf{b})\includegraphics[height = .3\textwidth,width = .3\textwidth]{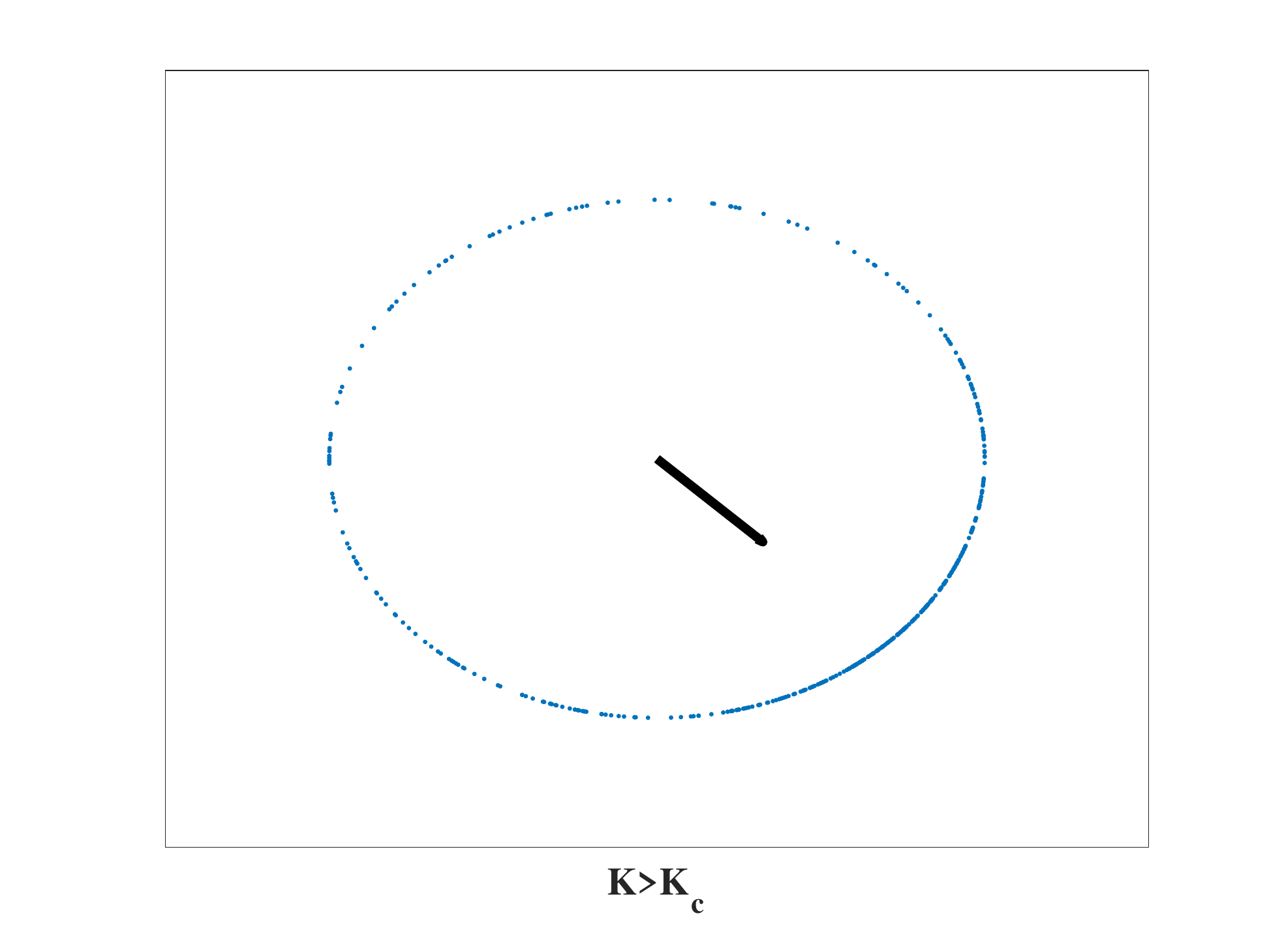}\;
\end{center}
 \lbl{f.1}
\caption{ The distribution of the oscillators in the phase space is shown for {\bf a})
$K<K_c$  and for {\bf b}) $K>K_c$. The complex order parameter is plotted as a black arrow.
It points to the center of mass of the coherence buildup.
}
\end{figure}
\begin{figure}
\begin{center}
\includegraphics[width = \textwidth]{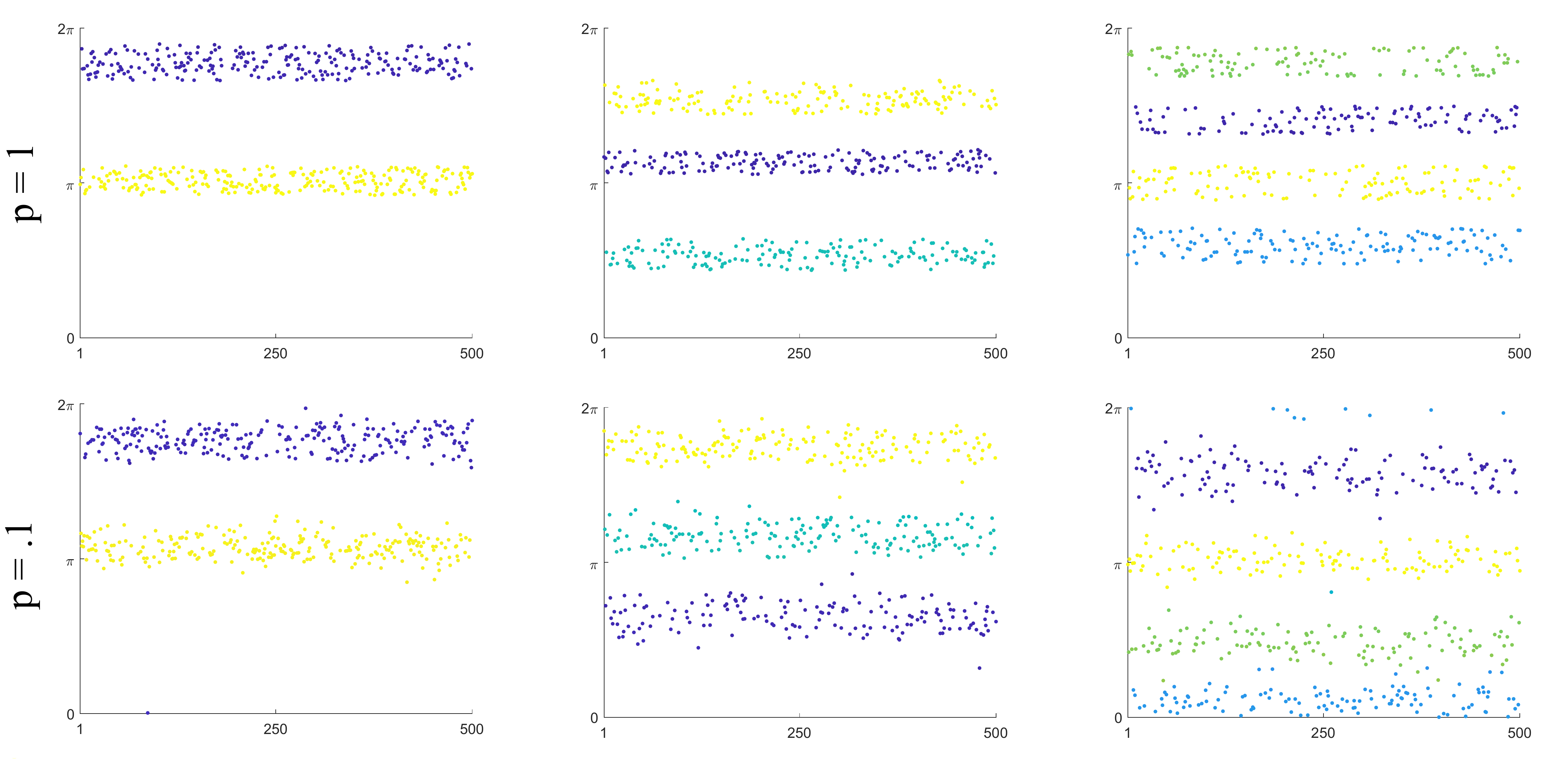}
\end{center}
\caption{ The  snapshots of coexisting distinct stable clusters in \eqref{2KMa}.
  The intrinsic frequencies are  chosen uniformly
  from $[-2,2]$. The values of other parameters are $K=3$ and $\gamma = .1$.  The plots in the top line are made for the model on complete graphs and those in the bottom line -
  for that on an ER graph with $p=.1$. 
  Oscillators are color coded by instantaneous velocity to reveal clustering. By choosing carefully initial conditions for the two models, one can generate $2-, 3-, 4-$ and other cluster states. 
}
\lbl{f.2}
\end{figure}
Determining stability of clusters is a challenging problem. For the second order KM with identical oscillators it
has been studied \cite{BBB16}, where the problem was reduced to the analysis of the
damped pendulum equation. For the model with random intrinsic frequencies and random network
topologies, linear stability of synchronization was studied in \cite{TOS19}.
For the model with random intrinsic frequencies, stability of clusters has not been studied before.
We show that this is  a multiscale problem. At a microscopic
level, the formation of clusters requires a mechanism by which the oscillators 
within a cluster stay coherent, i.e.,  synchronization within a cluster. On the other hand, 
clusters have nontrivial (macroscopic) dynamics of their own.
Thus, in addition to synchronization, the stability of clusters depends on the stability of the
macroscopic group motion.

In this paper, we study stability of clusters in the model with random intrinsic frequencies
and ER random connectivity.
We restrict to two-cluster states and assume a bimodal distribution of intrinsic frequencies.
These assumptions are made to simplify the presentation. The same approach can be used
for studying patterns with three and more groups of coherent oscillators.
Likewise, the multimodality of the intrinsic frequency distribution is not necessary for cluster formation.
The same mechanism is responsible for the formation of clusters when intrinsic frequencies are
distributed uniformly (see Fig.~\ref{f.2}). However, in this case additional care is needed
to identify the clusters analytically. We do not address this issue in this paper.
Furthermore, the same formalism applies to
the KM on other random graphs \cite{Med19}.
We develop a general framework for studying 
clusters in large systems of coupled phase oscillators with randomly distributed parameters.
As in \cite{BBB16}, we write down a low-dimensional system describing the macroscopic (group)
dynamics of clusters. Further, we  derive a system of kinetic PDEs characterizing the stochastic
dynamics of fluctuations with each cluster. The PDE for each cluster incorporates the information
about the group motion as well as the fluctuations in other clusters. The low--dimensional equation
for the group dynamics and the system of PDEs for fluctuations contain all information determining
the stability of clusters.
The former system can be further reduced to the equation of damped
pendulum and analyzed using standard methods of the qualitative theory of ordinary differential
equations \cite{Andronov-Theory}. On the other hand, the analysis of the two coupled Vlasov equations
is a hard problem, which we do not pursue in general. Instead,
we focus on parameter regimes when the two PDEs decouple, which simplifies
the analysis. The stability analysis in these parameter regimes suggests a scenario for the loss of
stability of a two-cluster state due to the loss of coherence in one of the clusters.
Specifically, we show that decoupling of the system of Vlasov equations results in the fluctuations in
one cluster being practically  independent from the fluctuations
in the other cluster. Thus, by controlling the fluctuations in one of the clusters
we can make it incoherent, while keeping the other cluster coherent. This provides a new scenario
of the loss of stability of a two-cluster state leading to the creation of a chimera state. 

The outline of the paper is as follows. In Section~\ref{sec.macro}, we develop a macro-micro decomposition
of the cluster dynamics into a low dimensional (group) motion of the centers of mass of two clusters
and the system of equations governing the fluctuations in each group. For the latter system, we derive
a system of two Vlasov PDEs describing the probability densities for the fluctuations in the limit as the
number of oscillators in each cluster tends to infinity. The macro-micro decomposition of the cluster
dynamics is the main tool and the main contribution of this paper. In Section~\ref{sec.group}, we
review the key facts about the dynamics of a damped pendulum \cite{Andronov-Theory} that will be needed
below. In Section~\ref{sec.chimera}, we turn to the analysis of fluctuations. We identify two parameter
regimes when the two Vlasov equations decouple and the coherence in each cluster can be analyzed
separately. We use linear stability analysis of the incoherent state in the KM with inertia \cite{Chiba-notes},
to locate the critical values for the loss of coherence in each cluster. Then we identify parameters
where oscillators in one cluster lose coherence, while the oscillators in the other
cluster remain synchronized. This leads to formation of  chimera states.
We illustrate this scenario with numerical experiments. Numerics are consistent
with the theoretical predictions. We conclude with a brief discussion of the main results in Section~\ref{sec.discuss}.

\begin{figure}
	\begin{center}
	\includegraphics[width = .7\textwidth]{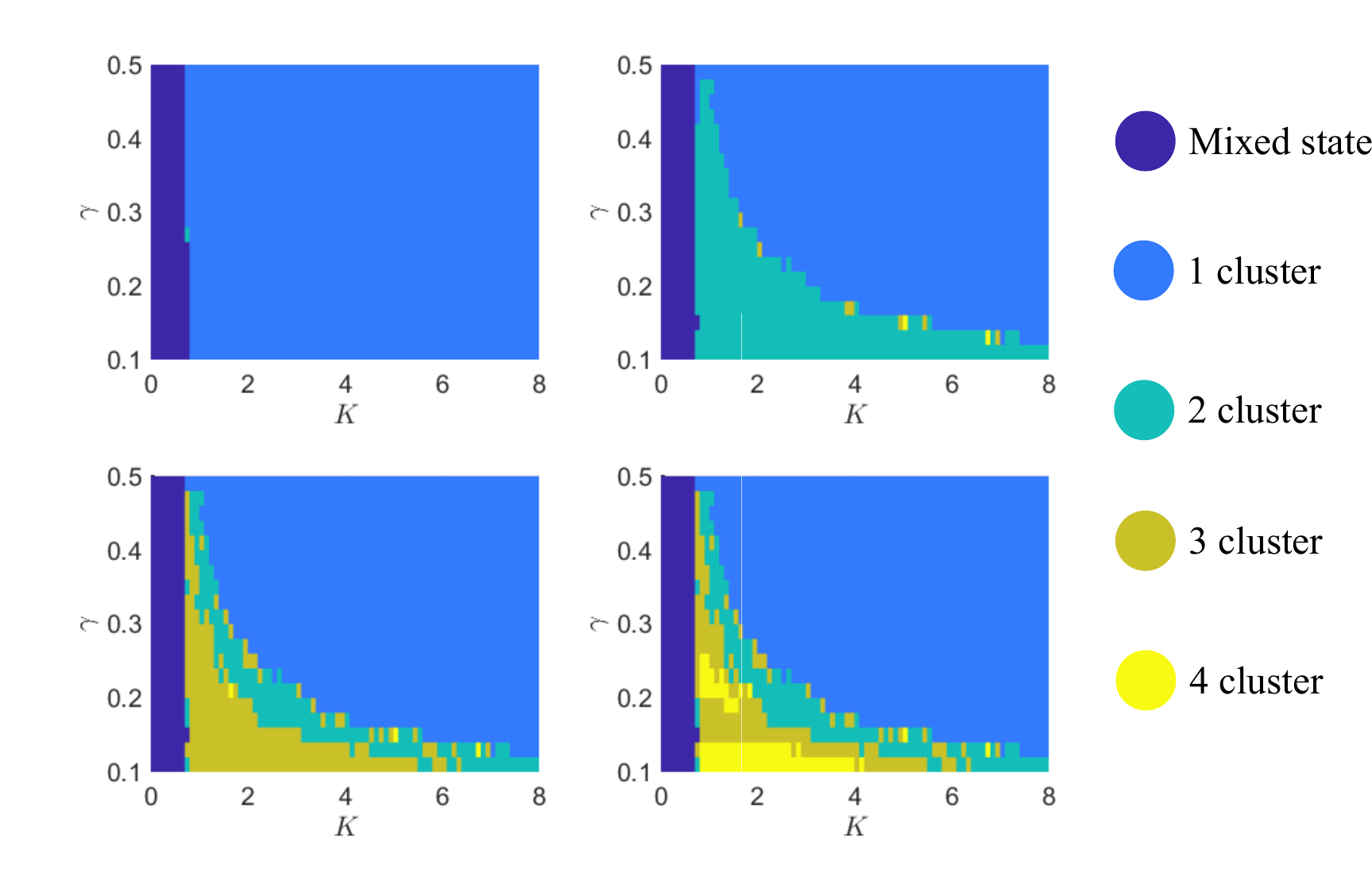}
	\end{center}
        \caption{Regions of existence of stable $d$-cluster states shown for 
        	 for $d=1$, $1\le d\le 2$,   $1\le d\le 3$ and $1\le d\le 4$ superimposed on each other. 
        	The diagrams show a substantial region in the
          parameter space with coexisting stable $1-$, $2-$, $3-$, and $4-$cluster states. The frequencies are
          sampled from the uniform distribution on
          $[-.5,.5].$}
\lbl{f.3}
\end{figure}

\section{The macro-micro decomposition}\lbl{sec.macro}
\setcounter{equation}{0}
\subsection{The model} 
For simplicity, we restrict our study to a two--cluster case\footnote{
  It is easy to generalize the equations determining stability of $d$-cluster stattes
  for $d\ge 2$, but the analysis of this system is already challenging for $d=2$.}.
To this end, we assume a bimodal distribution for 
$\omega_{n,i}$'s. Specifically, we assume that there are two groups of oscillators $u_1, u_2, \dots, u_m$ and 
$u_{m+1}, u_{m+1}, \dots, u_{m+l},$ $n=m+l$.
The intrinsic frequencies assigned to the oscillators
in the first and second groups are taken from probability distributions with densities 
$\tilde g_1(\omega)$ and $\tilde g_2(\omega)$ respectively. Denote the first two central moments by
\be\lbl{moments}
\bar\omega_1=\int \omega \tilde g_1(\omega)d\omega, \;  
\bar\omega_2=\int \omega \tilde g_2(\omega)d\omega, \; 
\sigma_1^2 =\int (\omega-\bar\omega_1)^2 \tilde g_1(\omega) d\omega,\;
\sigma_2^2 =\int (\omega-\bar\omega_1)^2 \tilde g_1(\omega) d\omega.
\ee
We assume
\be\lbl{assume-moments}
\delta=\bar\omega_2-\bar\omega_1>0,\quad
0<\sigma_1^2, \sigma_2^2 \ll\delta,
\ee
and $g_{1,2}(y)=\tilde g_{1,2}(\bar\omega_{1,2}+y)$ are even
unimodal functions. Further, we assume that the initial positions and velocities for each cluster
$\{u_{n,k}(0)\}_{k=1}^m,$ $\{\dot u_{n,k}(0)\}_{k=1}^m,$
$\{u_{n,m+k}(0)\}_{k=1}^l,$ $\{\dot u_{n,m+k}(0)\}_{k=1}^l$, are sequences of independent identically
distributed (each sequence has its own distribution in general) random variables, which satisfy assumptions
of the Strong Law of Large Numbers.

In addition, we assume that the underlying network has sparse  ER connectivity:
\be\lbl{Pedge}
\P(a_{n,ij}=1)=p_n,
\ee
where $(p_n)$ is a positive nonincreasing sequence that is either $p_n\equiv p\in [0,1]$ or 
$p_n\searrow 0$ such that $p_nn\to \infty$ as $n\to\infty$. In the latter case, we obtain a
sequence of sparse ER graphs of unbounded degree.
Thus, below we study the following system of ODEs\footnote{See
  \cite{Med19} for more details on the KM on sparse graphs.}:
\be\lbl{2KM}
\ddot u_{n,i} +\gamma \dot u_{n,i}=\omega_{n,i} +K(p_nn)^{-1}
\sum_{j=1}^n a_{n,ij} \sin\left(u_{n,j}-u_{n,i}+\alpha\right),\quad i\in [n].
\ee
The analysis of this can be easily generalized to a more general
W-random graph model (cf.~ \cite{Med19}). 
We restrict to the ER case to keep the notation simple.

%
%

\subsection{The group dynamics}
In this and the following subsections, we decompose the dynamics of clusters into two systems:
one governing the macroscopic dynamics of individual clusters and the second 
governing the microscopic dynamics of individual particles inside each cluster.
The former is a system of low dimensional ODEs and the latter is a system
of PDEs of Vlasov type. 

Denote
\be\lbl{anz}
U_{n,1}=m^{-1}\sum_{k=1}^m u_{n,k},\;
U_{n,2}=l^{-1}\sum_{k=1}^l u_{n,m+k},
\ee
where
\be\lbl{def-v}
\begin{array}{ll}
v_{n,i}=u_{n,i}-U_{n,1}, & \quad i\in [m],\\
v_{n,m+j}=u_{n,m+j}-U_{n,2}, & \quad j\in [l].
\end{array}
\ee
We assume that the dynamics in each cluster are (predominantly) coherent:
\be\lbl{assume:coherence}
\max_{i\in[n]} |v_i|\le \varepsilon\ll 1.
\ee

Adding up the first $m$ equations in \eqref{KM} and dividing by $m$, we have
\be\lbl{U1}
\begin{split}
\ddot U_{n,1} +\gamma\dot U_{n,1}
 =&\bar\omega_{n,1}+{K\over mnp_n}\sum_{i=1}^m\sum_{j=1}^m a_{n,ij} \sin\left(v_{n,j}-v_{n,i}\right) \\
+ &{K\over mnp_n}\sum_{i=1}^m\sum_{j=1}^l a_{n,ij}
\sin\left(U_{n,2}-U_{n,1}+[v_{n,m+j}-v_{n,i}]+\alpha\right).
\end{split}
\ee
Rewrite the last sum on the right--hand side of \eqref{U1}
as 
\be\lbl{sin-cos}
\begin{split}
\sum_{i=1}^m\sum_{j=1}^l 
a_{n,ij} \sin\left(U_{n,2}-U_{n,1}-\alpha+v_{n,m+j}-v_{n,i}\right) &=\sum_{i=1}^m\sum_{j=1}^l 
a_{n,ij}\left\{
\sin\left(U_{n,2}-U_{n,1}+\alpha \right) \cos\left( v_{n,m+j}-v_{n,i} \right)\right.\\
& +\left. \cos\left(U_{n,2}-U_{n,1}-\alpha\right)\sin\left(v_{n,m+j}-v_{n,i} \right)\right\}
\end{split}
\ee
and note
$$
\cos\left( v_{n,m+j}-v_{n,i} \right)=1-o(1),\; \sin\left(v_{n,m+j}-v_{n,i} \right)=o(1).
$$

After plugging \eqref{sin-cos} into \eqref{U1} and separating $O(1)$ terms,
we obtain the following IVP for the dynamics of the first cluster
\begin{eqnarray}\lbl{cl1}
\ddot U_{n,1} +\gamma\dot U_{n,1}& =&\bar\omega_{n,1} +{Kl\over n} \sin\left( U_{n,2}-U_{n,1}+\alpha\right),\\
\lbl{cl1-ic}
U_{n,1}(0) & =&m^{-1}\sum_{k=1}^m u_{n,k}(0),\\
\dot{U}_{n,1}(0) &=&  m^{-1} \sum_{k=1}^m \dot{u}_{n,k}(0)
\end{eqnarray}
where we also used
$$
m^{-1}\sum_{j=1} a_{n,ij}=p_n + o(1) \quad \mbox{with high probability}.
$$

We will assume that $mn^{-1}\rightarrow \chi\in (0,1)$, and so $ln^{-1}\rightarrow 1-\chi$
as $n\to\infty$.
By the Law of Large Numbers, $\bar\omega_{n,1}\to\bar\omega_1$, 
$m^{-1}\sum_{k=1}^m u_{n,k}(0)\to\bar u_1.$
Likewise, $m^{-1}\sum_{k=1}^m \dot{u}_{n,k}(0) \to \bar{\dot u}_1$
as $n\to\infty.$ Thus, for $n\gg 1$, \eqref{cl1}, \eqref{cl1-ic} is approximated by
\begin{eqnarray}\lbl{clst-1}
\ddot U_{1} +\gamma\dot U_{1}& =&\bar\omega_{1} +K(1-\chi) \sin\left( U_{2}-U_{1}+\alpha\right),\\
\lbl{clst-1-ic}
U_{1}(0) & =& \bar u_1, \\
\dot{U}_1(0) &=& \bar{\dot u}_1.
\end{eqnarray}

Similarly, we obtain the system approximating the dynamics of the second cluster
\begin{eqnarray}\lbl{clst-2}
\ddot U_{2} +\gamma\dot U_{2}& =&\bar\omega_{2} +K\chi \sin\left( U_{1}-U_{2}+\alpha\right),\\
\lbl{clst-1-ic}
U_{2}(0) & =&\bar u_2, \\
\dot{U}_2(0)&=& \bar{\dot u}_2.
\end{eqnarray}

\subsection{The fluctuations}
Next, we turn to the analysis of the fluctuations $v_{n,i}, \; i\in [n]$. After plugging in \eqref{anz} into the 
equation for the
oscillator~$i \in [m]$ and 
using \eqref{U1}, we have
\be\lbl{corr-1}
\begin{split}
\ddot v_{n,i} +\gamma\dot v_{n,i} &=
\xi_{n,i} +{K\over np_n} \sum_{j=1}^m a_{n,ij}
\sin\left( v_{n,j}-v_{n,i} +\alpha\right) \\
&+ {K\over n p_n}\sum_{j=1}^l a_{n,ij}\left\{\sin\left(U_{n,2}-U_{n,1}+\alpha\right) \left[
    \cos\left(v_{n,m+j}-v_{n,i}\right)-1\right] \right.\\
  & +\left.
\cos\left(U_{n,2}-U_{n,1}+\alpha\right) \sin\left(v_{n,m+j}-v_{n,i}\right) \right\} \\
&-{K\over mnp_n} \sum_{i=1}^m \sum_{j=1}^l a_{n,ij}\sin\left(U_{n,2}-U_{n,1}\right)
\left[
\cos\left(v_{n,m+j}-v_{n,i}\right)-1\right] \\
&-{K\over mn p_n} \sum_{i=1}^m \sum_{j=1}^l a_{n,ij}\cos\left(U_{n,2}-U_{n,1}+\alpha\right)
\sin\left(v_{n,m+j}-v_{n,i}\right)
\; i\in [m],
\end{split}
\ee
where $\xi_{n,i}=\omega_{n,i}-\bar\omega_{n,1}.$  For large $n$,  $\xi_{n,i}, i\in [n]$ are approximated by 
iid RVs $\xi_i, \; i\in [n],$ having probability density $g_1$.

Since $|v_{n,i}|=o(1)$, terms
$$
1-\cos\left(v_{n,m+j}-v_{n,i}\right)=2\sin\left({v_{n,m+j}-v_{n,i}\over
    2}\right)^2, \; j\in [l],
$$
are of higher order and can be dropped.  Further, we approximate $U_{n,1}$ and $U_{n,2}$ by
$U_1$ and $U_2$ respectively. Thus, \eqref{corr-1} simplifies to
\be\lbl{corr-1a}
\begin{split}
\ddot v_{n,i} +\gamma\dot v_{n,i} &=
\xi_i +{K\over np_n} \sum_{j=1}^m a_{n,ij}
\sin\left( v_{n,j}-v_{n,i} +\alpha\right) \\
&+ {K\over np_n}\sum_{j=1}^l a_{n,ij} 
\cos\left(U_{2}-U_{1}+\alpha\right) \sin\left(v_{n,m+j}-v_{n,i}\right) \\
&-{K\over mn p_n} \sum_{i=1}^m \sum_{j=1}^l a_{n,ij}\cos\left(U_2-U_1+\alpha\right)
\sin\left(v_{n,m+j}-v_{n,i}\right)
\; i\in [m],
\end{split}
\ee

Next, we show that 
\be\lbl{double-sum}
{1\over mn p_n} \sum_{i=1}^m \sum_{j=1}^l
a_{n,ij}  \sin\left(v_{n,m+j}-v_{n,i}\right) =o(1).
\ee 

By the Taylor's formula and triangle inequality, we have
\be\lbl{taylor-expand}
\begin{split}
\left| {1\over mn p_n} \sum_{i=1}^m \sum_{j=1}^l
a_{n,ij}  \sin\left(v_{n,m+j}-v_{n,i}\right) \right| & \le 
\left| {1\over mnp_n} \sum_{i=1}^m \sum_{j=1}^l
a_{n,ij} \left(v_{n,m+j}-v_{n,i}\right)
\right|\\
& +
\left|{1\over mnp_n} \sum_{i=1}^m \sum_{j=1}^l
a_{n,ij} \left(v_{m+j}-v_i\right)^3\right|\\
&=
\left| {1\over mn} \sum_{i=1}^m \sum_{j=1}^l {a_{n,ij}\over p_n} \left(v_{n,m+j}-v_{n,i}\right) 
\right| + O(\epsilon^3).
\end{split}
\ee
Further, since 
$$
\sum_{j=1}^m v_{n,j}=\sum_{j=1}^l v_{n,m+j}=0,
$$
the sum in first term on the right hand side of \eqref{taylor-expand} can be written as
$$
{1\over mn} \sum_{i=1}^m \sum_{j=1}^l
{a_{n,ij}\over p_n} \left(v_{n,m+j}-v_{n,i}\right)= {1\over mn}
\sum_{i=1}^m \sum_{j=1}^l \xi_{n,ij} \left(v_{n,m+j}-v_{n,i}\right),
$$
where $\xi_{n,ij}={a_{n,ij}\over p_n}-1$ are independent zero--mean
random variables. If we assume that all $v_i$'s are bounded almost
surely, then the application of Bernstein inequality yields that for
any $0<\varepsilon<1/2$
\be\lbl{first-term}
{1\over mn} \sum_{i=1}^m \sum_{j=1}^l
{a_{n,ij}\over p_n} \left(v_{n,m+j}-v_{n,i}\right)= {1\over mn}
\sum_{i=1}^m \sum_{j=1}^l \xi_{n,ij} \left(v_{n,m+j}-v_{n,i}\right) =
O(n^{-{1\over 2}+\varepsilon})
\ee
with high probability. The combination of \eqref{taylor-expand} and
\eqref{first-term} yields \eqref{double-sum}.

Thus, we arrive at the following equation 
\be\lbl{corr-1-final}
\begin{split}
\ddot v_{n,i} +\gamma\dot v_{n,i} &=
\xi_i^{(1)} +{K\over n p_n} \sum_{j=1}^m a_{n,ij}
\sin\left( v_{n,j}-v_{n,i}+\alpha \right) \\
&+ {K\over np_n} c(t)\sum_{j=1}^l 
a_{n,ij}\sin\left(v_{n,m+j}-v_{n,i}\right),
\; i\in [m],
\end{split}
\ee
where
\be\lbl{coupling}
c(t)=\cos\left(U_2-U_1+\alpha\right).
\ee
The terms on the first line of \eqref{corr-1-final} constitute the KM for one 
cluster. The sum on the second line yields the contribution from the other cluster.

Similarly, we derive the system of equations of fluctuations in the second cluster
\be\lbl{corr-2}
\begin{split}
\ddot v_{n,m+i} +\gamma\dot v_{n,m+i} &=
\xi^{(2)}_i +{K\over n} \sum_{j=1}^l 
\sin\left( v_{n,m+j}-v_{n,m+i} +\alpha \right) \\
&+  
{K\over n} c(t) \sum_{j=1}^m \sin\left(v_{n,j}-v_{n,m+i}+\alpha\right),
\quad i\in [l],
\end{split}
\ee
where $\xi^{(2)}_i,\; i\in [l]$ are iid RVs whose distribution has density $g_{\omega_2}$.

To analyze large systems \eqref{corr-1} and \eqref{corr-2} we use the mean field limit approximation.
To this end, suppose $f_1(t,u,v,\omega)$ and $f_2(t,u,v,\omega)$ stand for 
the probability densities of the oscillators in the first and second clusters respectively.
Then
\be\lbl{MF-1}
\p_t f_1 + \p_{u} \left( v f_1\right) + \p_{v} \left( V_1 f\right) =0
\ee
where
\be\lbl{V-1}
\begin{split}
V_1(u,v,\omega)&:= \omega -\gamma v +K\chi \int_{\T\times \R\times\R}
\sin\left( \phi- u+\alpha\right) f_1(t,\phi, \psi,\lambda) g_1(\lambda) 
d\phi d\psi d\lambda\\
&+
 K(1-\chi) c(t) \int_{\T\times \R\times\R}
\sin\left( \phi- u+\alpha\right) f_2(t,\phi, \psi,\lambda) 
g_{\omega_1}(\lambda) d\phi d\psi d\lambda,\quad \chi=mn^{-1}.
\end{split}
\ee

Similarly,
\be\lbl{MF-2}
\p_t f_2 + \p_{u} \left( v f_2\right) + \p_{v} \left( V_2 f_2\right) =0
\ee
where
\be\lbl{V-2}
\begin{split}
V_2(u,v,\omega)&:= \omega -\gamma v +K(1-\chi) \int_{\T\times \R\times\R}
\sin\left( \phi- u+\alpha\right) f_2(t,\phi, \psi,\omega) g_2(\lambda) d\phi d\psi d\lambda\\
&+
 K \chi c(t) \int_{\T\times \R\times\R}
\sin\left( \phi- u+\alpha\right) f_1(t,\phi, \psi,\omega) g_1(\lambda) d\phi  d\psi d\lambda.
\end{split}
\ee

In the numerical experiments below, we are going to use the following order parameters computed
for each cluster:
\be\lbl{order}
R_1(t)={1\over m} \sum_{j=1}^m e^{\1 u_{n,j}},\quad 
R_2(t)={1\over l} \sum_{j=1}^l e^{\1 u_{n,j}}.
\ee
The modulus of $R_1 (R_2)$ measures the degree of coherence in cluster~1 (2): with values close to $0$
corresponding to a high degree of mixing and those close to $1$ corresponding to a high degree of coherence.


\section{The damped pendulum equation}\label{sec.group}

To continue we need to understand the group dynamics \eqref{clst-1}, \eqref{clst-2}.
To this end, we change variables to
\be\lbl{new-var}
X=U_2-U_1, \quad S=(1-\chi)^{-1} U_1+\chi^{-1} U_2, 
\ee
and rewrite \eqref{clst-1}, \eqref{clst-2} as 
\begin{eqnarray}\lbl{X-eqn}
\ddot X +\gamma\dot X &=&\delta -K\left(\chi \sin(X-\alpha)+(1-\chi)\sin(X+\alpha)\right),\quad 
\delta:=\bar\omega_2-\bar\omega_1>0,\\
\lbl{Y-eqn}
\ddot S +\gamma\dot S &=& \delta_0,\quad 
\delta_0:= (1-\chi)^{-1}\bar\omega_1+\chi^{-1}\bar\omega_2.
\end{eqnarray}

In the remainder of this section, we restrict to $\chi = 1/2$, as this is the value used in all our experiments.
For the treatment of \eqref{X-eqn} for other values of $\chi$, we refer the interested reader to
\cite{BBB16}. For $\chi=1/2$, we have 
\begin{equation}
	\ddot X +\gamma\dot X = \delta-K\cos(\alpha)\sin(X).
\end{equation}


Equation \eqref{X-eqn} is the damped pendulum equation with constant torque. Qualitative dynamics
of \eqref{X-eqn} can be understood using phase plane analysis \cite{Andronov-Theory}.
To this end,  rewrite \eqref{X-eqn} as
\begin{eqnarray}\lbl{cons-X}
\dot X &=&  Y,\\
\lbl{cons-Y}
\dot Y &=& \delta -\gamma Y-K\cos\alpha\,\sin X.
\end{eqnarray}

Note that by rescaling variables and parameters  $Y:=\delta^{-1/2}Y,$ $\gamma:=\delta^{-1/2}\gamma$,
and $K:=\delta^{-1/2}K\cos\alpha$,
and changing time we can scale out $\delta$:
\begin{eqnarray}\lbl{pendX}
\dot X &=&  Y,\\
\lbl{pendY}
\dot Y &=& 1 -\gamma Y-K \sin X.
\end{eqnarray}
Thus, without loss of generality one can set $\delta=1.$ 

We summarize the phase plane analysis of the damped pendulum equation \eqref{pendX}, \eqref{pendY} and 
refer the interested reader to \cite{Andronov-Theory} for more details. First, it is easy to see that for $K>1$ the
system has a pair of equilibria:
\be\lbl{equilib}
(X_e,0) \;\mbox{and}\; (\pi-X_e, 0), \; X_e=\arcsin K^{-1}.
\ee
The former is a stable focus while the latter is a saddle. They collide in a saddle-node bifurcation at $K=1$ and disappear 
for $K<1$. Further, for $K<1$ the Poincar\'e-Bendixson theorem implies existence of a limit cycle, which must be stable as
the divergence of the vector field is equal to $-\gamma<0$ (Fig.~\ref{f.pplane}\textbf{a}).
The limit cycle persists for $K\ge 1$ provided that
$0<\gamma\le \gamma_{hom}(K)$ (Fig.~\ref{f.pplane}\textbf{b}). At $\gamma=\gamma_{hom}(K)$, the system undegoes a homoclinic 
bifurcation (Fig.~\ref{f.pplane}\textbf{c}) . Thus, there are three parameter regimes with qualitatively distinct dynamics shown in Fig.~\ref{f.bifD}:
In (I) and (II) the attractor is a limit cycle and stable focus respectively. In (III) both the limit cycle and the stable focus
coexist.

   \begin{figure}
     \centering
                     \textbf{a}) \includegraphics[width = .3\textwidth]{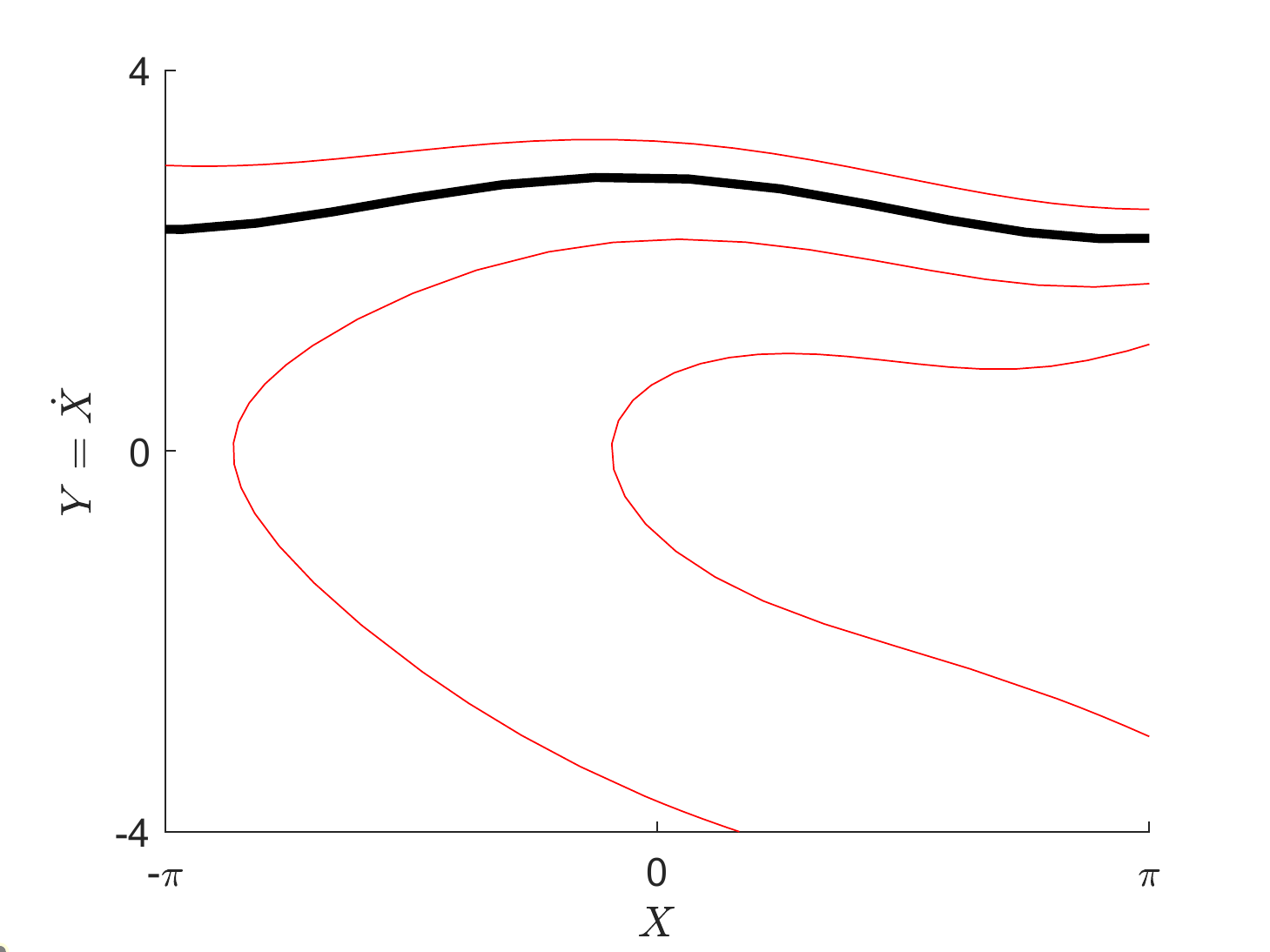}
                     \textbf{b})   \includegraphics[width = .3\textwidth]{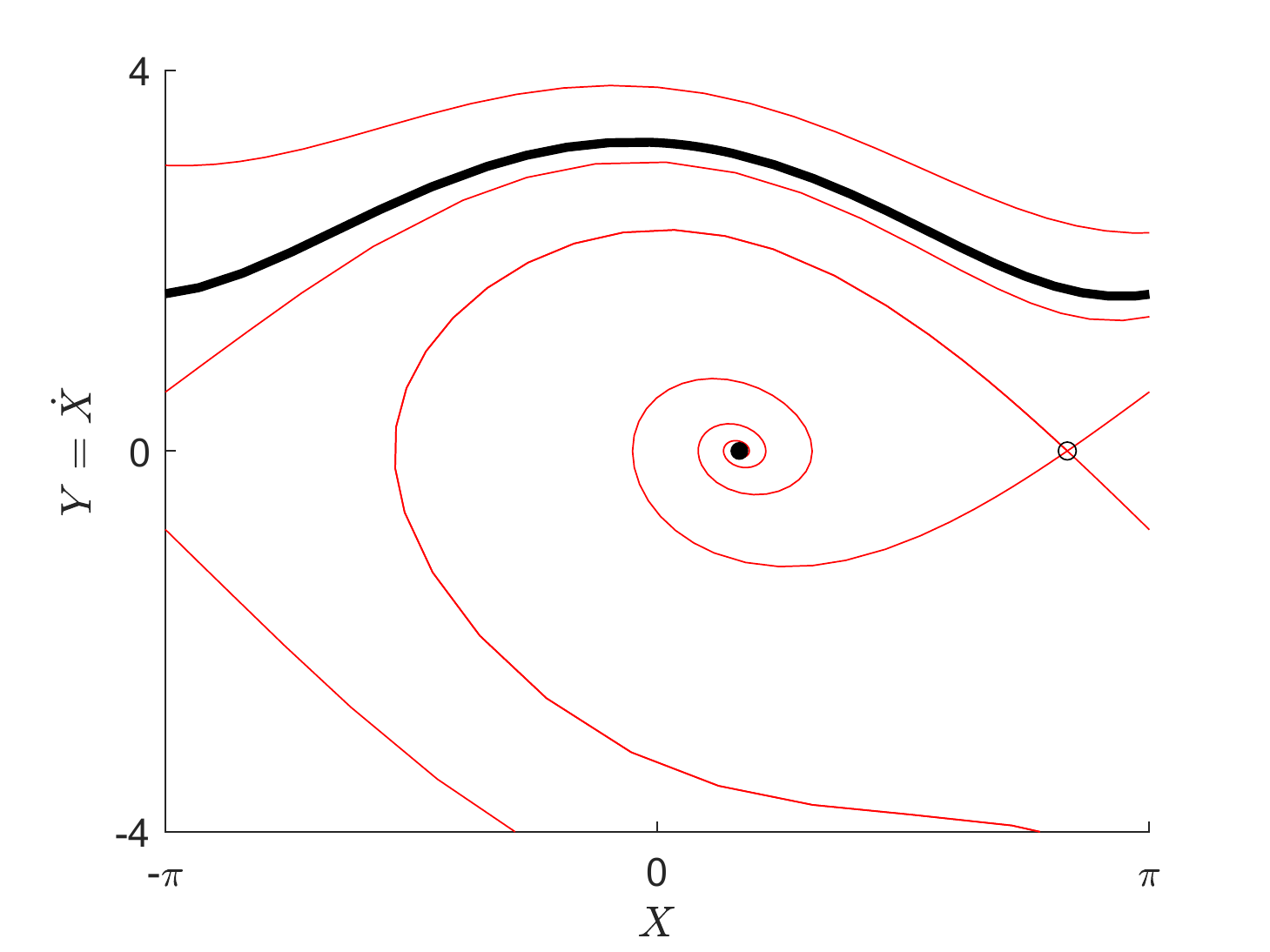}
                     \textbf{c}) \includegraphics[width = .3\textwidth]{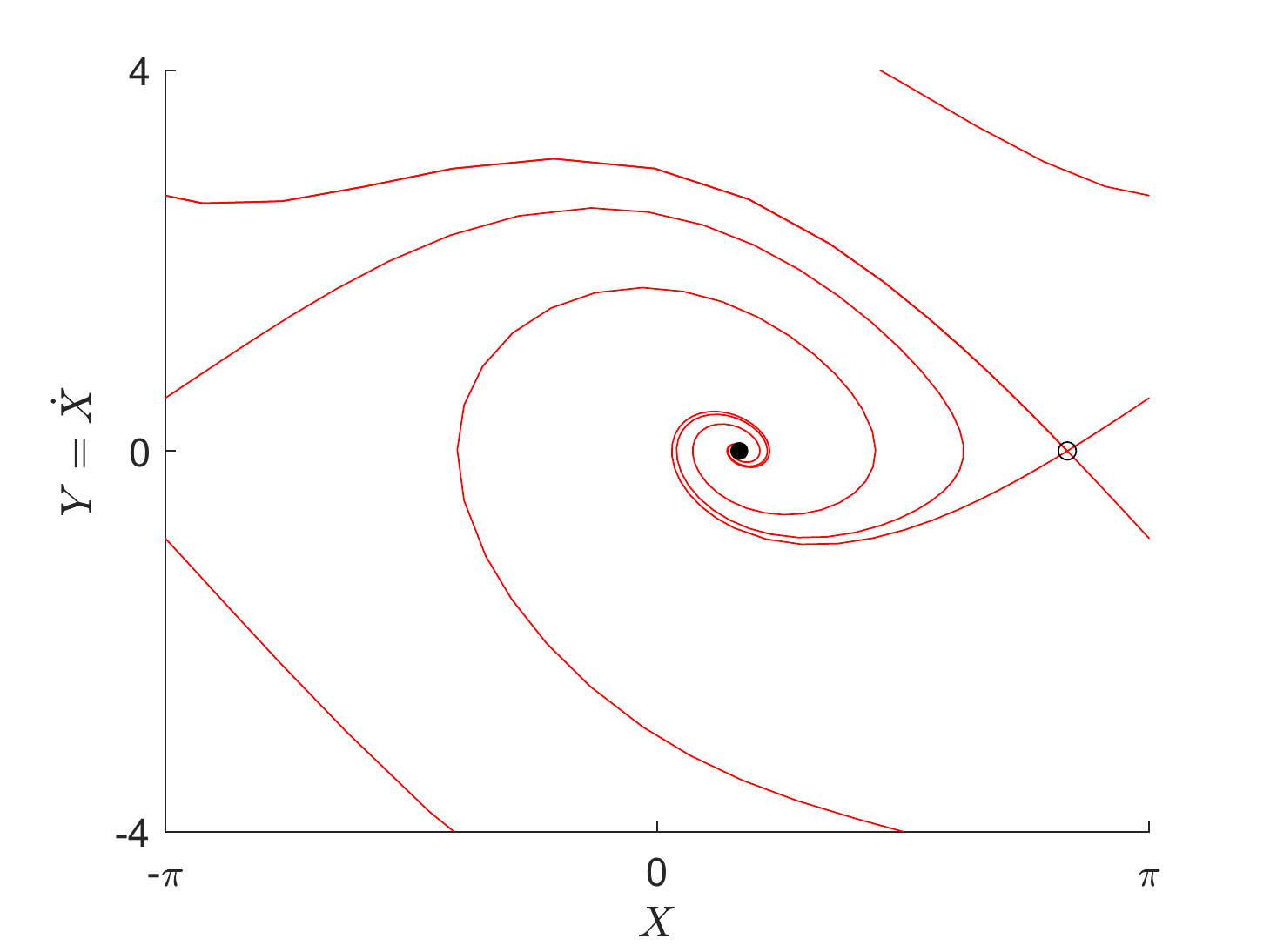}\\
            \centering
                \caption{Phase plane plots illustrating three structurally stable regimes for the damped pendulum.
                 \textbf{a}) The black trajectory corresponds to the stable periodic orbit (when viewed on the cylinder).
                  \textbf{b}) Two fixed points appear in a saddle--node bifurcation. Thus, we have a stable focus coexisting
                  with a stable periodic orbit. \textbf{c}) The periodic orbit disappears in a homoclinic bifurcation.
                  The stable focus remains the only attractor.
                }\lbl{f.pplane}
\end{figure}

\begin{figure}
	\centering
		\includegraphics[width = .3\textwidth]{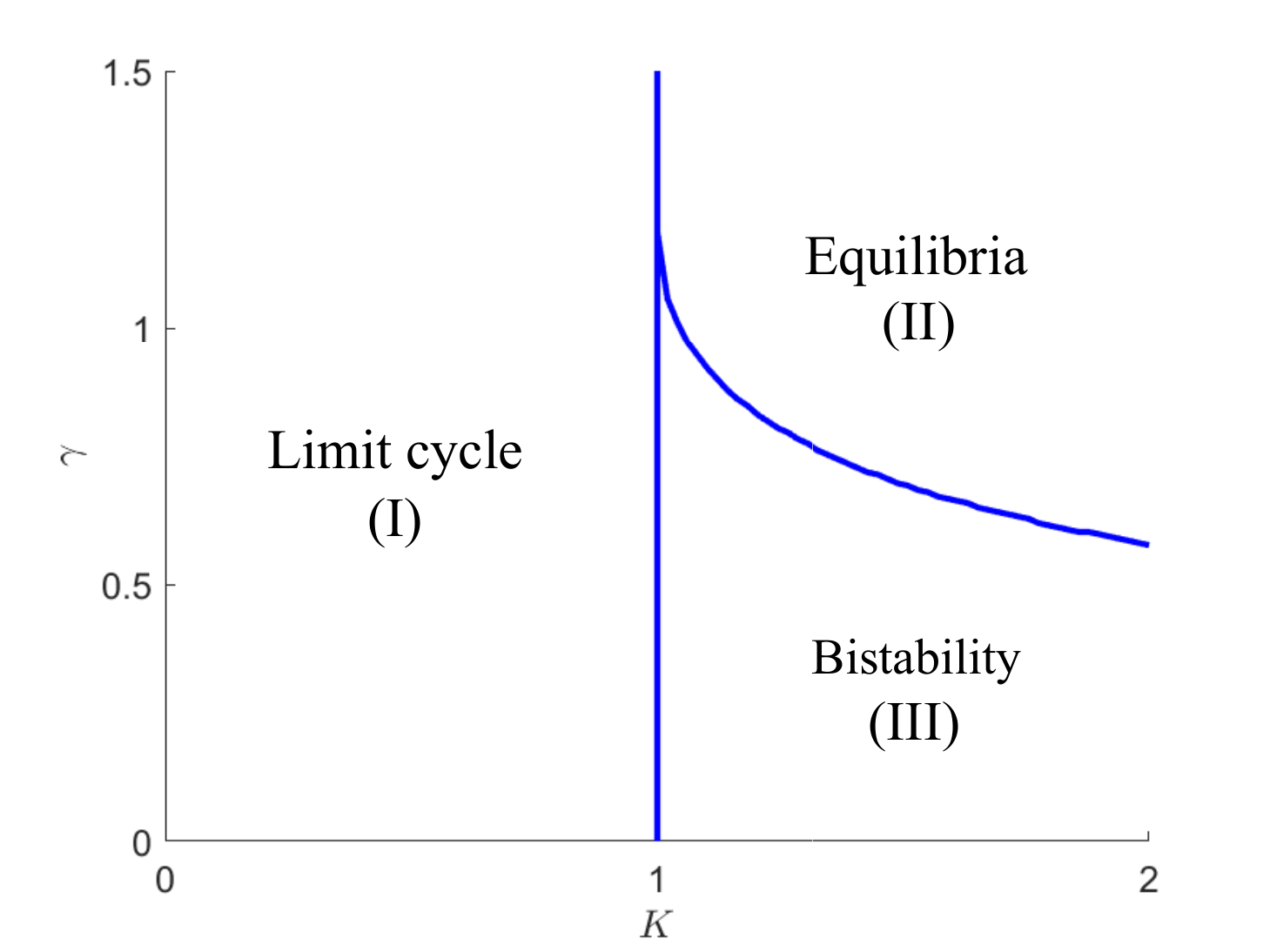}
  	\caption{Bifurcation diagram illustrating three qualitatively
          distinct regimes in the damped pendulum model \eqref{pendX}, \eqref{pendY}.}
        \lbl{f.bifD}
\end{figure}

\section{The loss of coherence and chimera states}\lbl{sec.chimera}
\setcounter{equation}{0}

\subsection{The overview of synchronization in the second--order KM}
In this section, we describe a mechanism for the loss of stability of a
two-cluster state due to the loss of synchronization in one of the clusters.
We show that this leads to the creation of chimera states.
To this end, it is instructive first to review synchronization in a single all--to--all coupled population of 
second--order phase oscillators:
\be\lbl{1-2KM}
\ddot u_{n,i}+\gamma\dot u_{n,i} =\xi_i +Kn^{-1} \sum_{j=1}^n \sin \left( u_{n,j}-u_{n,i}\right),
\quad
i\in [n],
\ee
where $\xi_i$ are IID RVs taken from a probability distribution with density $g$. Throughout this discussion,
we assume that $g$ is a unimodal even function.
If the initial conditions are drawn from the continuous
probability distribution then the distribution 
of the phase of oscillators in the extended phase space  $\SS\times \R\times\R$ remains absolutely continuous
 with respect to the Lebesgue measure for every $t>0$.
The density $f(t,u,v,\omega)$ satisfies the following Vlasov equation (cf.~\cite{ChiMed19a})
\be\lbl{1-MF}
  \p_t f + \p_{u} \left( v f \right) + \p_{v} \left( V f\right) =0,
\ee
where
\be\lbl{1-V}
V:= \omega -\gamma v +K \int_{\T\times \R\times \R}
\sin\left( \phi- u\right) f (t,\phi, \psi,\omega) g(\lambda) d\phi d\psi d\lambda.
\ee
The Vlasov equation \eqref{1-MF}, \eqref{1-V} has a steady state solution:
\be\lbl{MFsteady}
\bar f(u,v)={\delta_{\omega/\gamma}(v)\over 2\pi}.
\ee
It describes the configuration when phases are distributed uniformly over the unit circle, while velocities
are localized around $\omega/\gamma$. This is an incoherent or mixing state. Linear stability analysis of \eqref{1-MF}
about $\bar f$ shows that there is a critical value $K_c>0$ such that the mixing state is stable for $K\in [0, K_c]$ and
and is unstable for $K>K_c$. For $\alpha=0$ the value of $K_c$ is known explicitly \cite{Chiba-notes} 
\be\lbl{Kc}
K_c=2\left(\pi g(0) -\int_\R {\gamma g(\gamma\omega)\over \gamma^2 +\omega^2}~d\omega\right)^{-1}.
\ee

\subsection{The loss of coherence within a cluster}

\begin{figure}[h]
	\centering
	{\bf a}) \includegraphics[height = .25\textwidth, width=.6\textwidth]{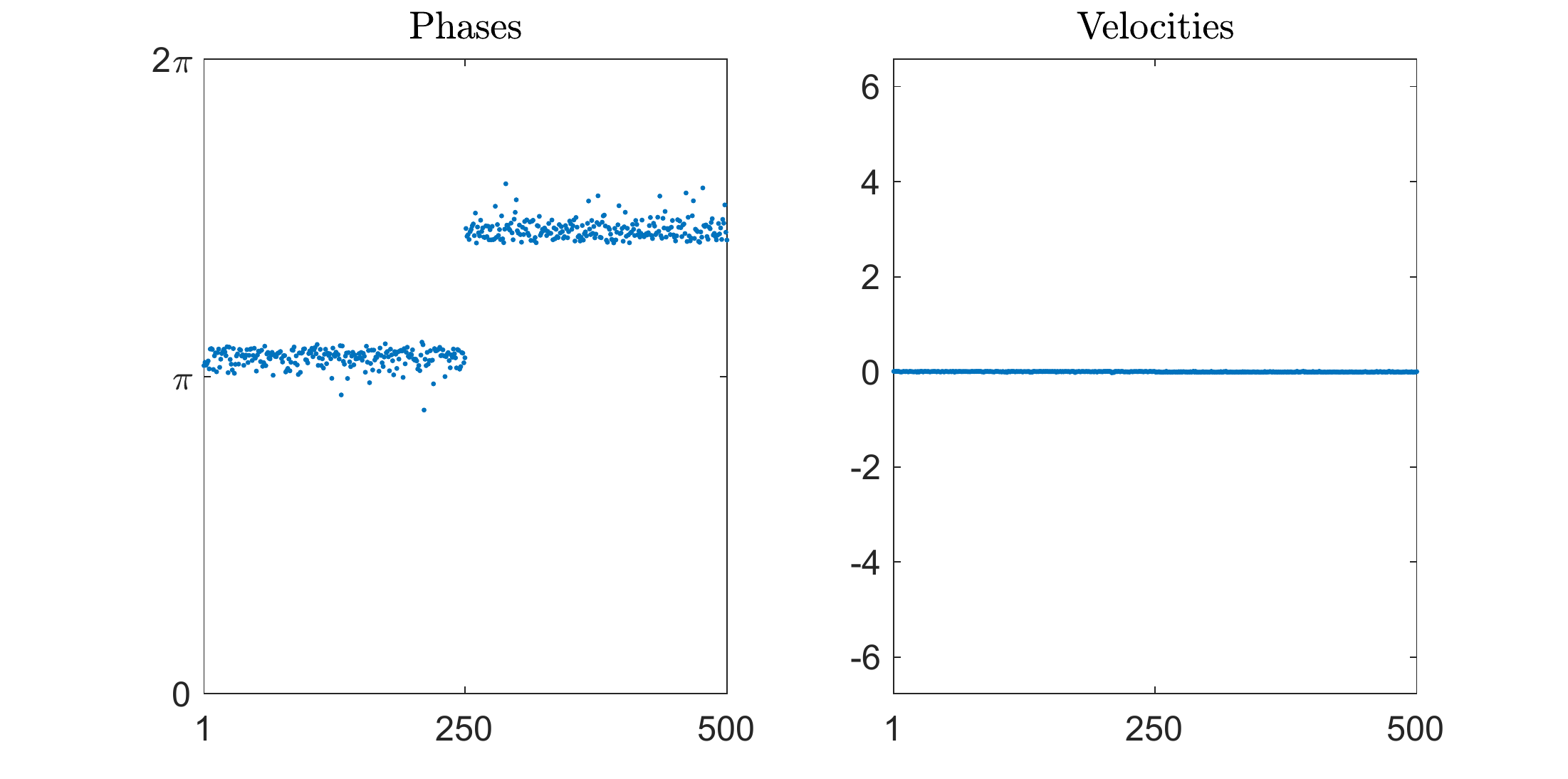}
	
	{\bf b}) 	\includegraphics[height = .25\textwidth,, width=.6\textwidth]{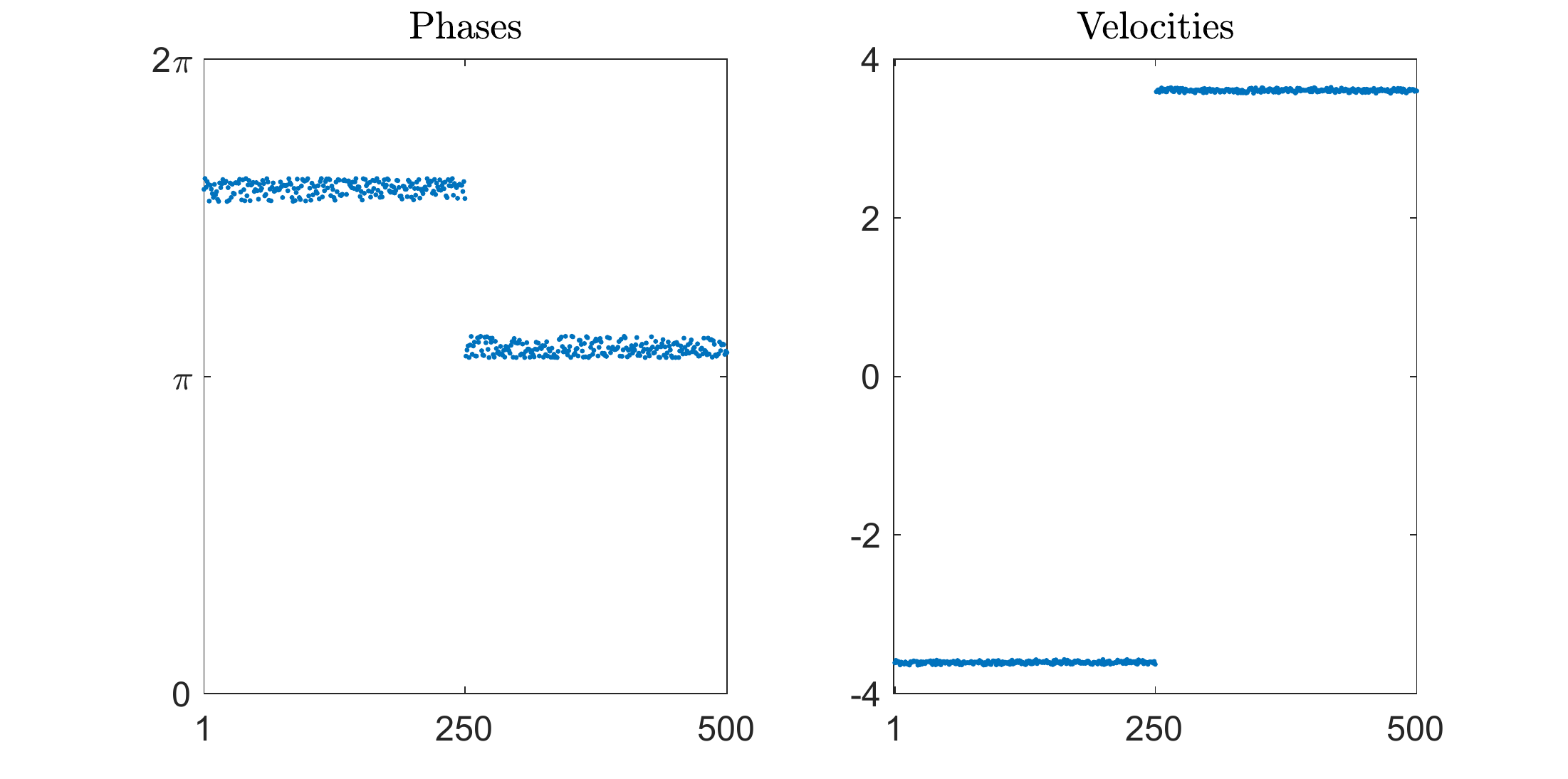}
	
	\caption{Two qualitatively distinct types of clusters: {\bf a}) stationary and {\bf b}) moving.
          The corresponding stable states of the pendulum equation \eqref{pendX}, \eqref{pendY} are
          a stable fixed point and a limit cycle respectively. The values of parameters used for both plots are 
          $\gamma = .1$ and $\omega$ chosen from $\mathcal{N}(\pm .5,.05)$.}
	\label{f.5}
\end{figure}

The macro--micro decomposition yields the following picture of cluster dynamics in the second order KM.
The macroscopic evolution of two subpopulations is described by the damped pendulum equation \eqref{pendX},
\eqref{pendY}. On the other hand the fluctuations in the two subpopulations are described by the system of two
coupled Vlasov equations \eqref{MF-1}, \eqref{V-1} and \eqref{MF-2}, \eqref{V-2}. The coupling between
\eqref{MF-1}, \eqref{V-1} and \eqref{MF-2}, \eqref{V-2} is modulated by the group dynamics through
$c(t)$ (see \eqref{coupling}). For stability of a two-cluster
configuration, we need a stable solution of the pendulum equation. In addition, 
we need fluctuations in both groups to remain small.
There are two qualitatively distinct stable states of the equation for the group
motion:
\begin{description}
  \item[A)] a stable fixed point resulting in the phase locked (stationary) clusters (Fig.~\ref{f.pplane} \textbf{b},\textbf{c}),
  \item[B)] a stable limit cycle resulting in two clusters moving in opposite directions (Fig.~\ref{f.pplane} \textbf{a},\textbf{b}).
  \end{description}
  The corresponding clusters are shown in Fig.~\ref{f.5}.
For each of this cases, we show that one can desynchronize the oscillators in one cluster without
affecting the oscillators in the other cluster.

We consider stationary clusters first.
Recall the damped pendulum equation \eqref{pendX}, \eqref{pendY} governing the group dynamics.
For $K>1$, it has a pair of fixed points (Fig.~\ref{f.bifD}), one of which is stable (cf.~\eqref{equilib}).
We suppose that the group dynamics is driven by the stable equilibrium. We will locate parameter
regimes where the fluctuations in the two clusters become practically independent. Then we demonstrate
that the fluctuations in each cluster can be controlled separately. In particular, we will desynchronize one
cluster, while keeping the other one coherent.

\begin{figure}
	\centering
	{\bf a}) \includegraphics[width=.4\textwidth]{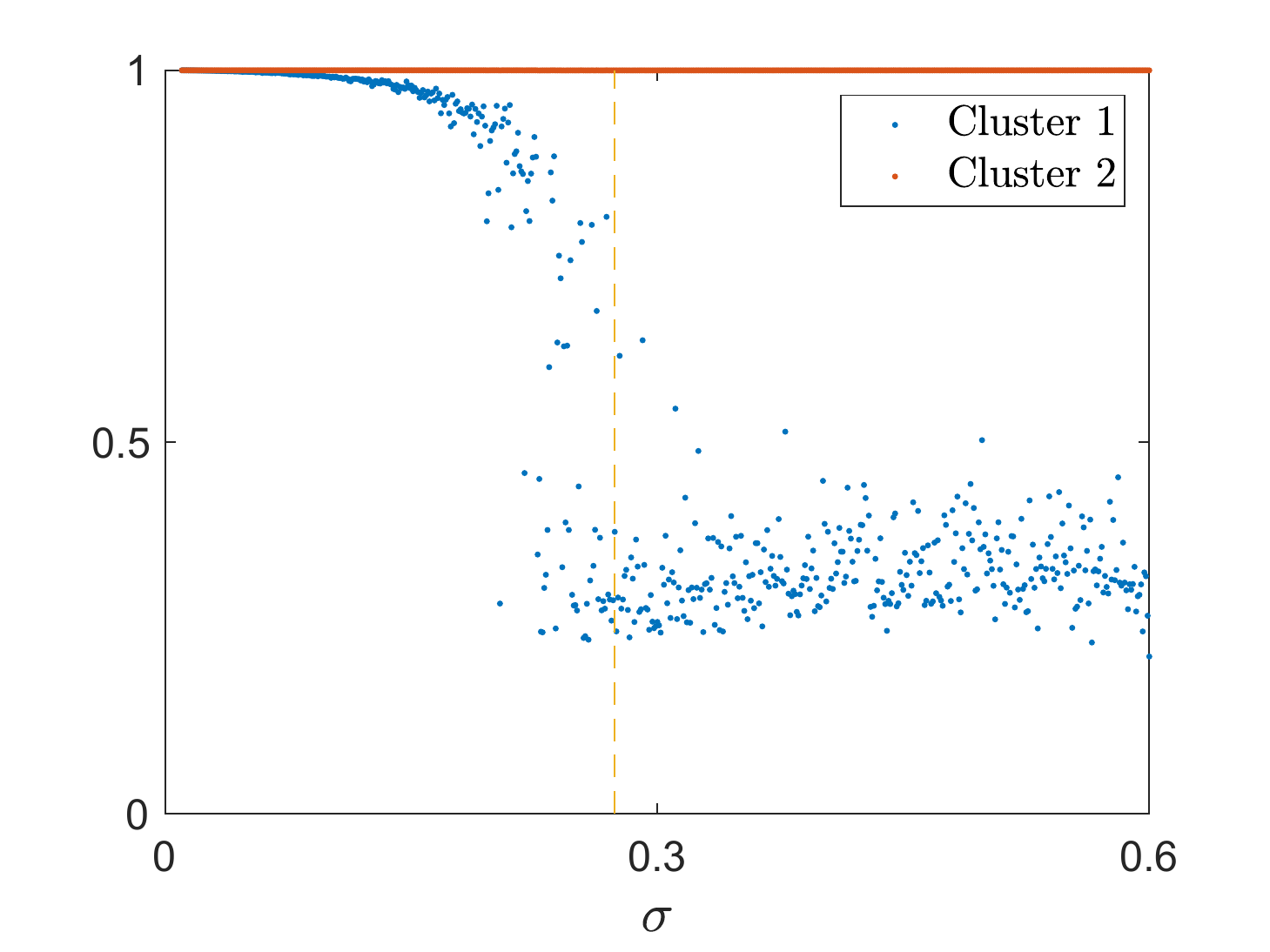}
	{\bf b})\includegraphics[width=.4\textwidth]{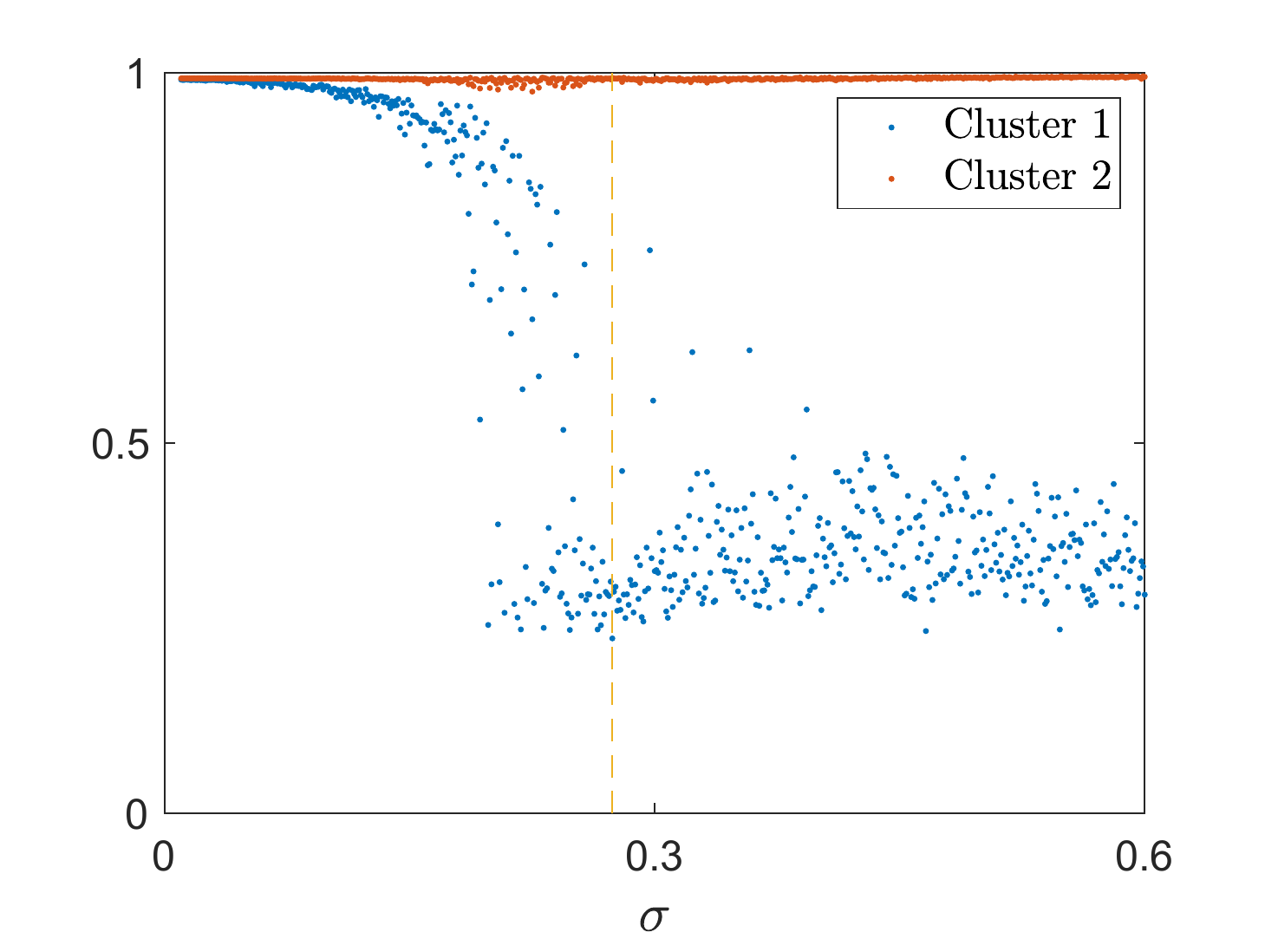}

    \hspace*{.4cm}\includegraphics[width = .35\textwidth]{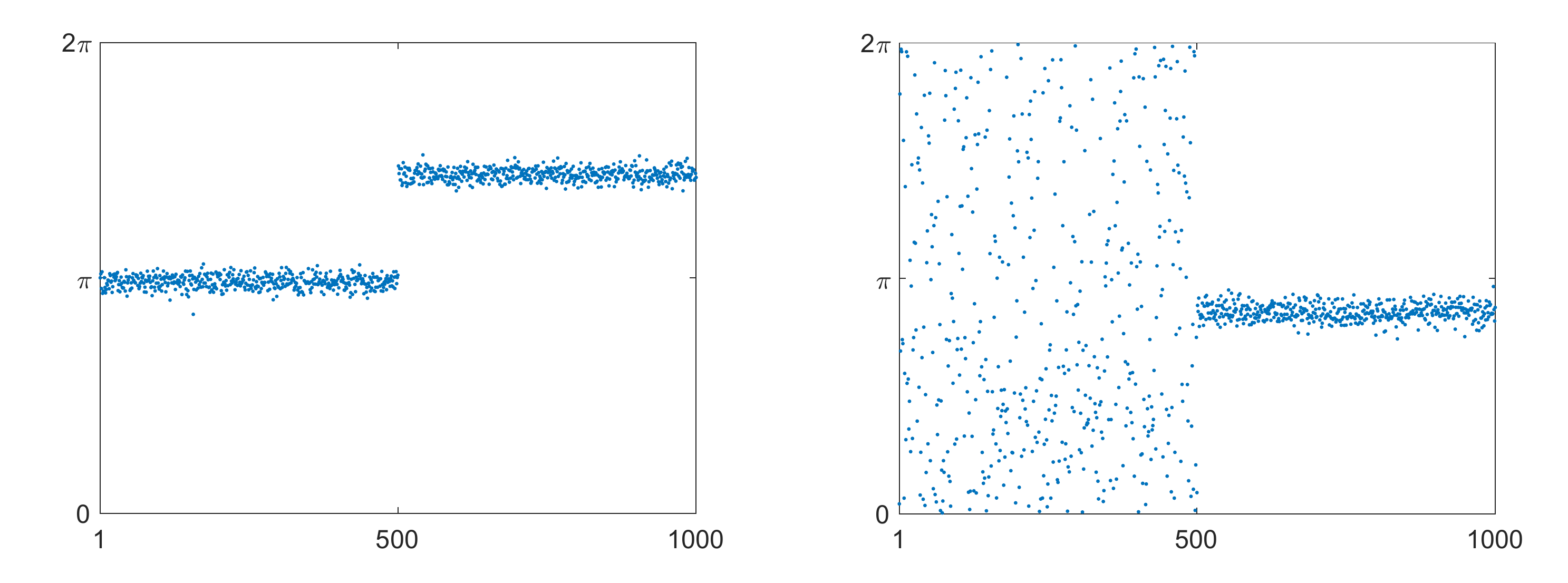}\hspace*{1.1cm}
    \includegraphics[width = .35\textwidth]{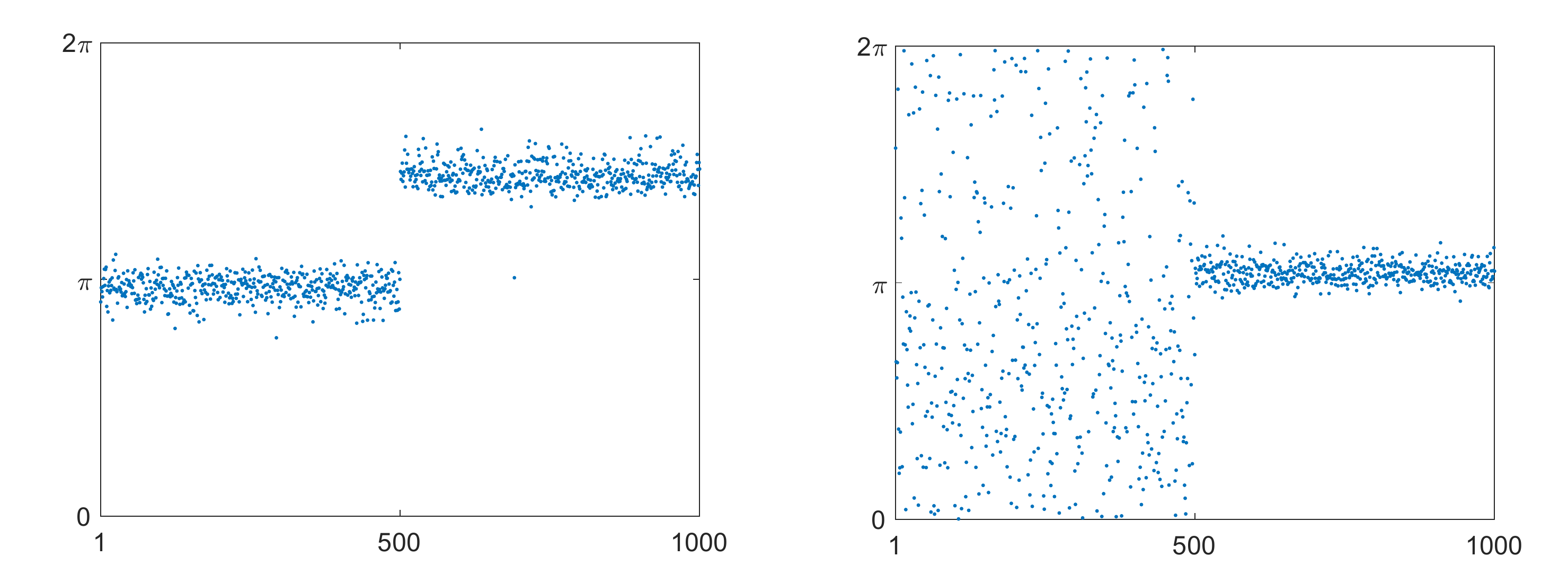}

\vspace*{1cm}
    {\bf c})\includegraphics[width=.4\textwidth]{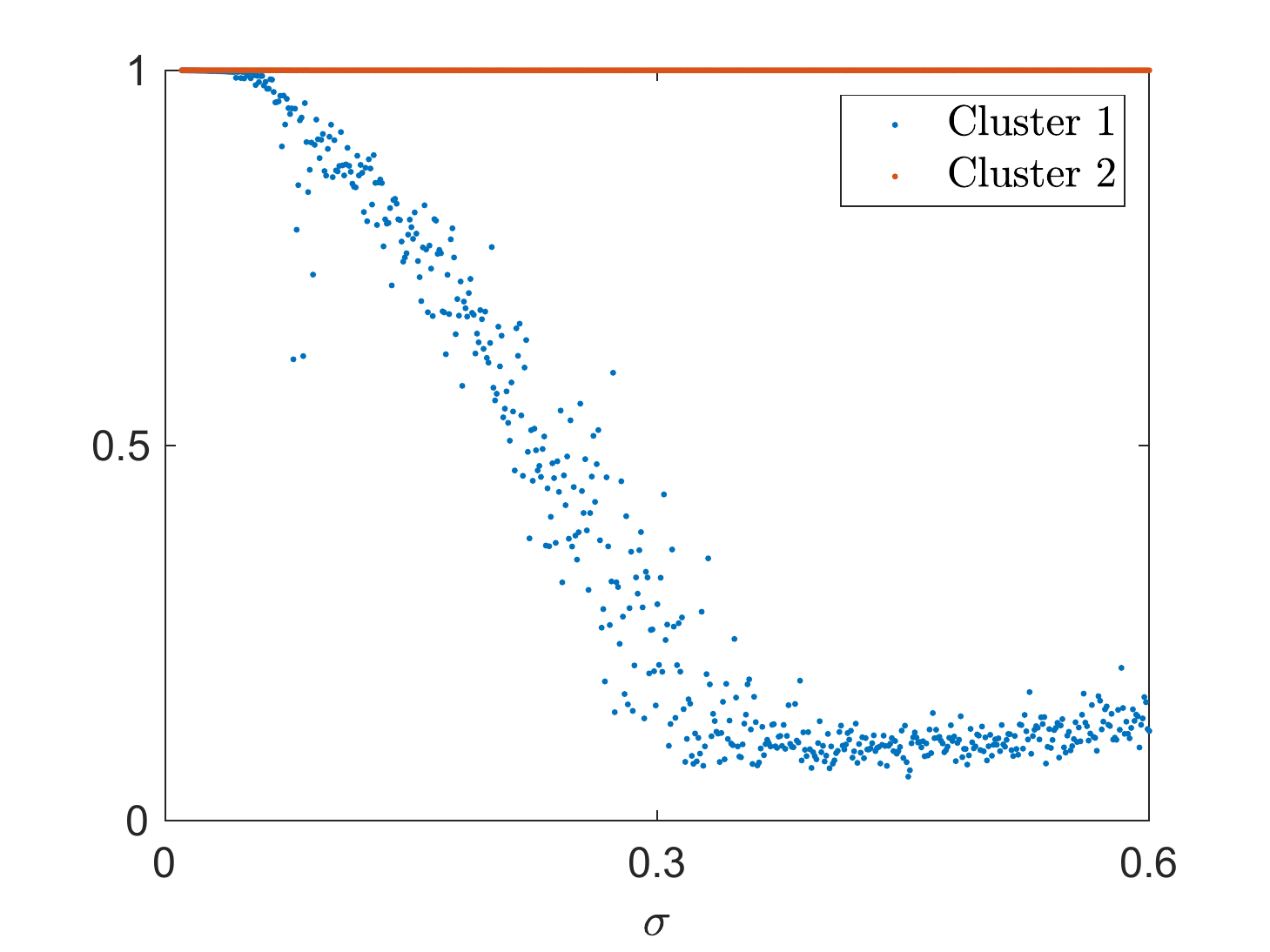}
	{\bf d})\includegraphics[width=.4\textwidth]{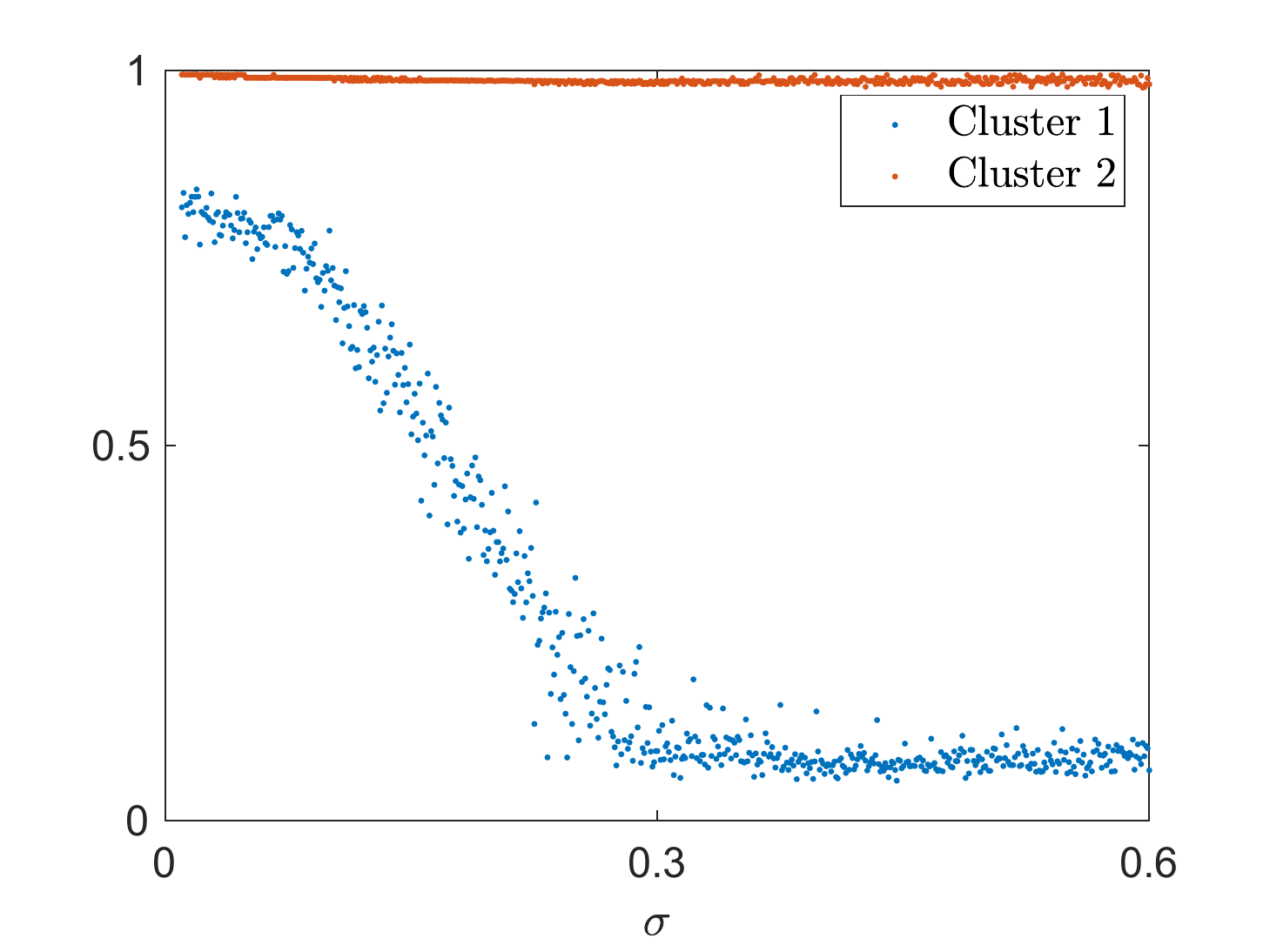}
	
	    \hspace*{.4cm}\includegraphics[width = .35\textwidth]{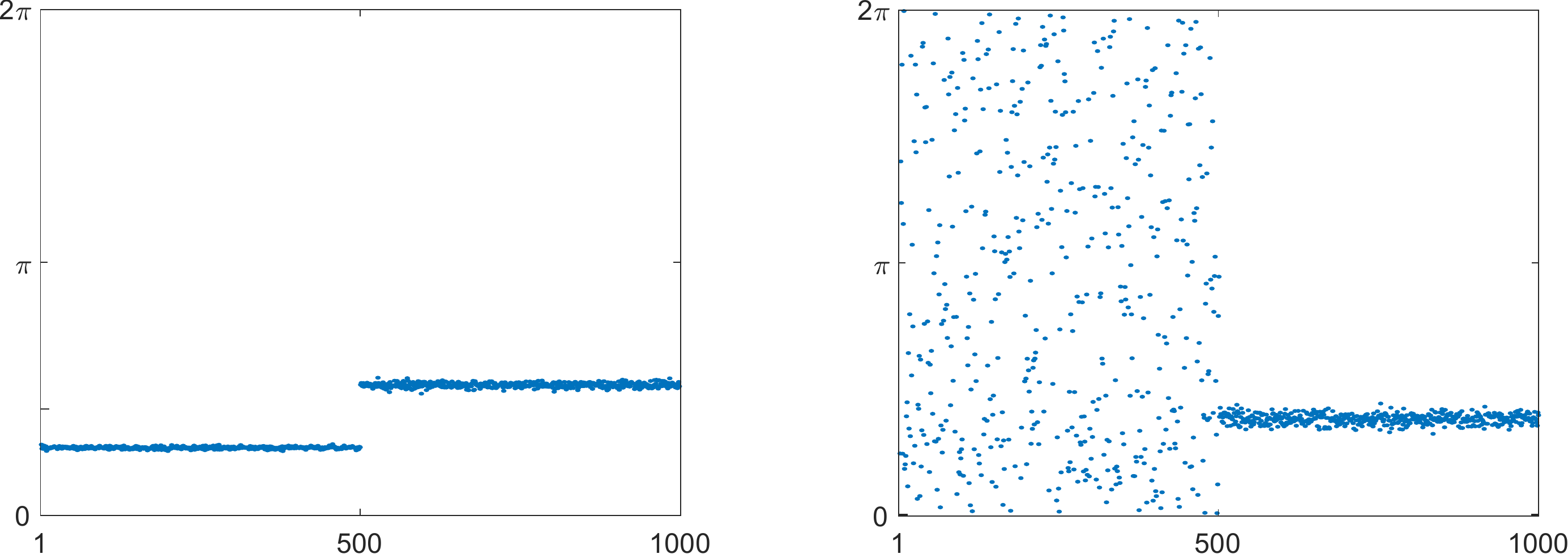}\hspace*{1.1cm}
	\includegraphics[width = .35\textwidth]{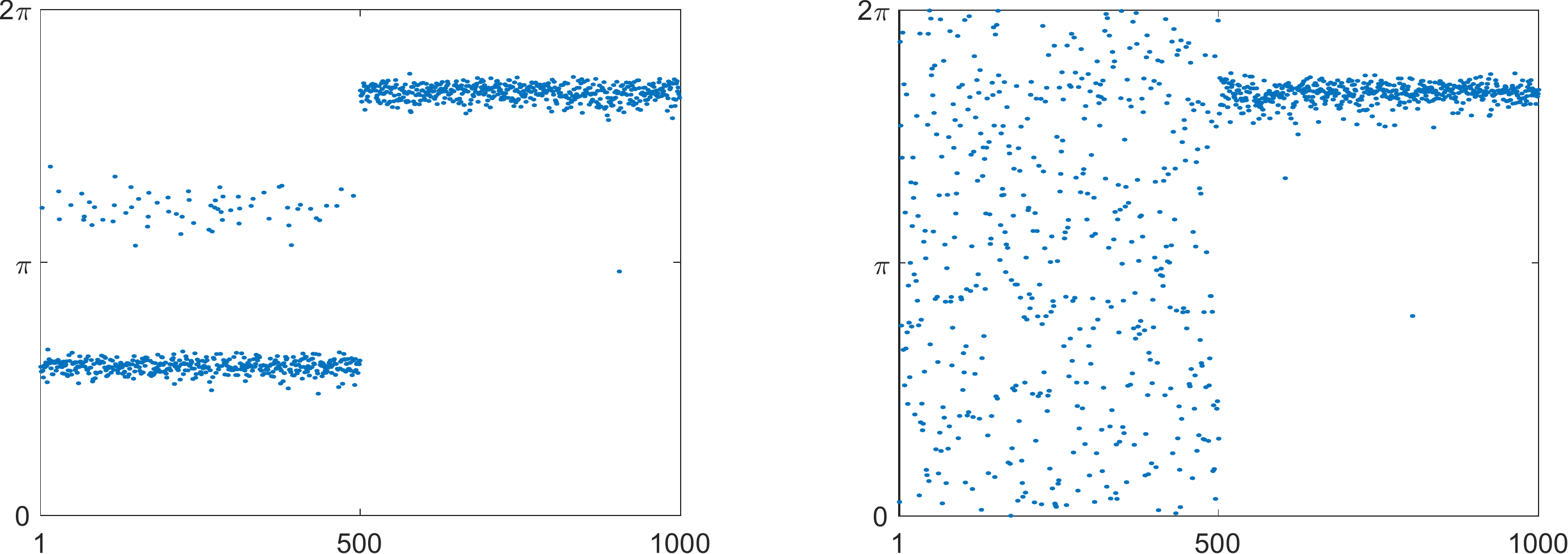}
	
	\caption{For $0<K-1\ll 1$ the system for the group dynamics \eqref{pendX}, \eqref{pendY}
          has a stable fixed point. When the group dynamics is driven by the stable fixed point, $c(t)\approx 0$
          (see the text). For a given $K$, we use \eqref{Kc} to compute the critical value of the variance
          $\sigma_*^2,$ at
          which the incoherent state loses stability.
          When the variance of the distribution of the intrinsic frequencies for the first cluster
          is increased beyond $\sigma^*$ (dotted line) the
          first cluster desynchronizes. This results in the formation of chimera state.
          {\bf a}) All-to-all coupling and  {\bf b}) ER connectivity with $p=.1$. Other parameters
          are $K=1.1$ and $\gamma =1$.
          The same experiment was repeated for $K=2$, $\gamma =.5$, and $\alpha = 1.0472$. The results are shown
          for  \textbf{c}) all-to-all coupling and \textbf{d}) ER graph with $p=.5$.}
	\label{f.6}
\end{figure}

We start with the case of $\alpha=0$. When the system \eqref{pendX}, \eqref{pendY} is at
the stable equilibrium (cf. \eqref{equilib}), 
\begin{equation}\lbl{U1-U2}
U_2- U_1 = \arcsin\left(\frac{1}{K\cos(\alpha)}\right),
\end{equation}
so $c(t) = \frac{\sqrt{K^2-1}}{K}$. Thus, for $K$ just above $1$,
\begin{equation}
0< K-1\ll 1,
 \end{equation}
 we have $c(t)\approx 0$. In this regime, the two Vlasov equations describing the coherence in the two
 clusters are practically decoupled. Thus, we can treat each cluster as a separate population of oscillators and
 compute the critical values of the coupling strength using \eqref{Kc} for each cluster separately.
 Next we choose the variances of the distributions of intrinsic frequencies $\sigma_1^2$ and $\sigma_2^2$
 such that
 $$
 K_c(\sigma_1^2)<K<K_c(\sigma_2^2).
 $$
 Then for a given value of $K$ the mixing state is stable for the first cluster, while it is unstable for the second
 cluster.
 As a result, we get a chimera state with the oscillators in the first cluster distributed uniformly while the
 oscillators
 in the second cluster remain synchronized (see Fig.~\ref{f.6}).

\begin{figure} 
	\centering
	       {\bf a}) \includegraphics[width = .45\textwidth]{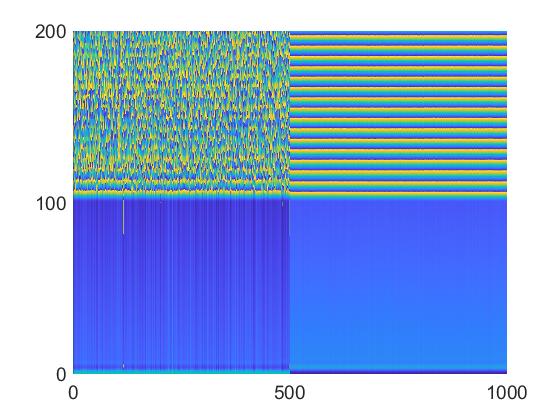}  
		{\bf b}) \includegraphics[width = .45\textwidth]{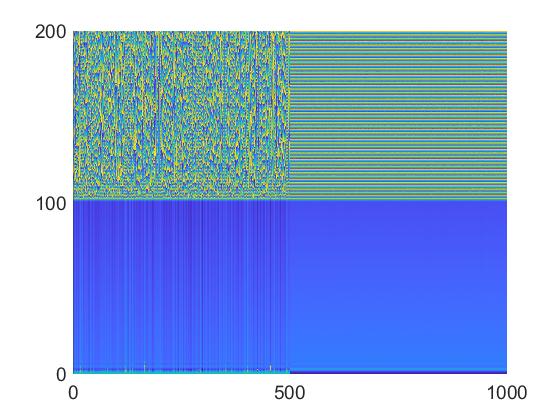}
		
		\caption{Plots show the evolution of oscillators
			(horizontal axis) over time  (vertical axis) with phase indicated by color. Before $t=100$, we fix $\alpha=0$. {\bf a}) At $t=100$ instantaneously let $\alpha = \alpha^*$ (the solution of \eqref{find_alpha}). See also Figure \ref{f.9}. {\bf b}) At $t=100$ let $\alpha$ evolve by \eqref{eq:alpha_ode}. See also Figure \ref{f.dotalpha}. In both cases the first cluster desynchronizes resulting in the emergence of a chimera state. Parameters are $\gamma=1$, $K=5$, and frequencies are chosen from
			$\mathcal{N}(-.5,.9)$ and $\mathcal{N}(.5,.05)$.}
		\label{f.chimera_phases}
\end{figure}

\begin{figure}
	\centering
    
	 {\bf a}) \includegraphics[width = .6\textwidth]{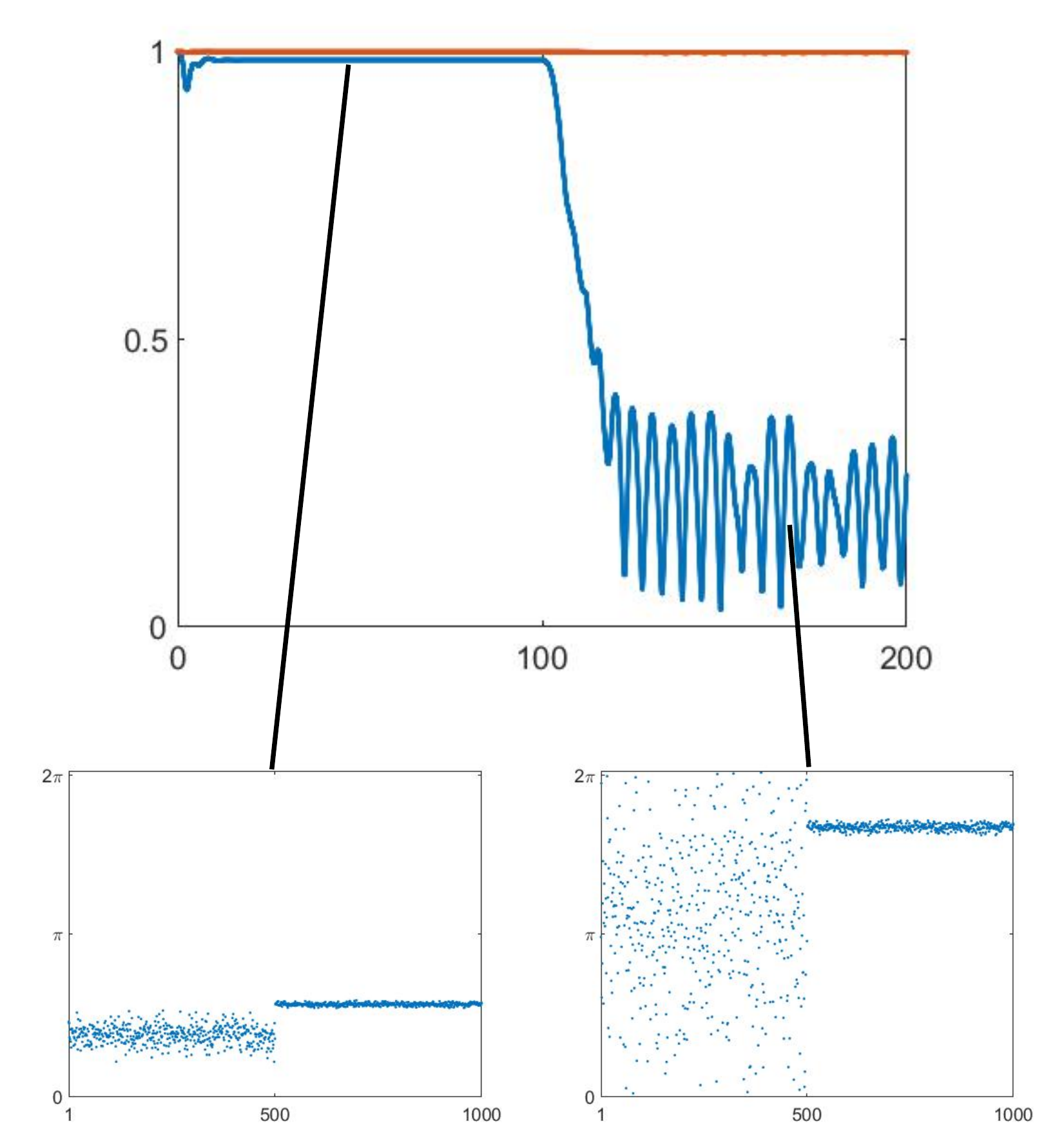}
	{\bf b}) \includegraphics[width  =.3\textwidth]{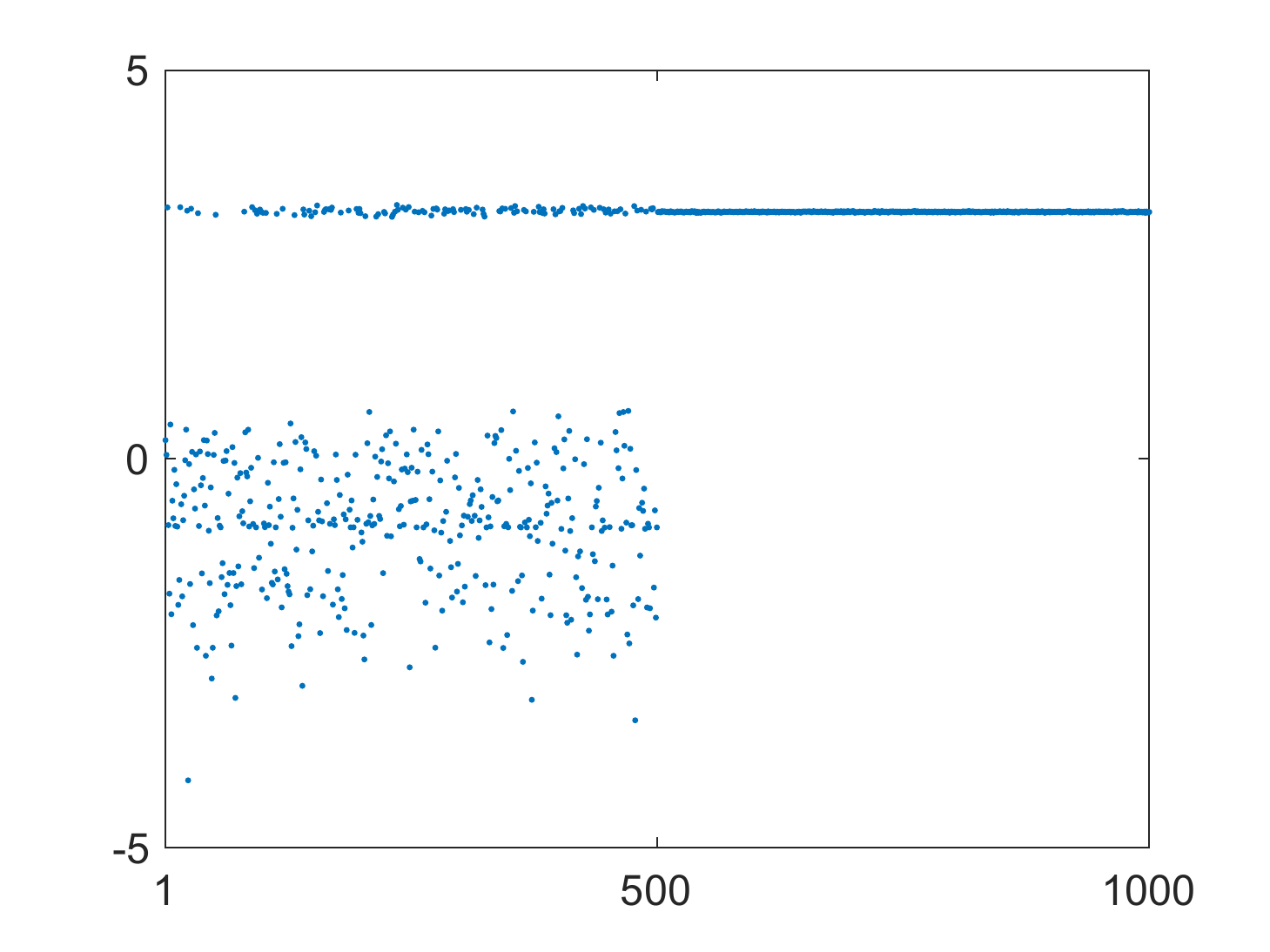}
	\caption{Starting the simulation with $\alpha=0$ fixed, we observe phase-locked solutions.	
          At $t=100$ we
          instantaneously let $\alpha = \alpha^*$ (the solution of \eqref{find_alpha}) and observe the
          emergence of a chimera state.  {\bf a}) Shows the order parameter of each
          cluster over time, together with snapshots showing both phase locked and chimera states. {\bf b}) Shows a snapshot of oscillator velocities from the chimera state. Here $\gamma=1$, $K=5$, and frequencies are chosen from
          $\mathcal{N}(-.5,.9)$ and $\mathcal{N}(.5,.05)$.}
	\label{f.9}
\end{figure}

The same idea can be used to generate chimera states for an arbitrary value of $K$ by changing $\alpha$.
In this case, $c(t)=\cos(U_2-U_1+\alpha)$ from \eqref{U1-U2} we have
\begin{equation*}
c(t) = \cos\left(\arcsin\left(\frac{1}{K\cos(\alpha)}\right)+\alpha\right).
\end{equation*}
Choosing $\alpha:=\alpha^\ast\in (0,\arccos(K^{-1}))$ such that
\begin{equation}\label{find_alpha}
\arcsin\left(\frac{1}{K\cos(\alpha^\ast)}\right)+\alpha^\ast = \frac{\pi}{2}.
\end{equation}
we can make $c(t)\equiv 0$.
With this choice of $\alpha$, the two Vlasov equations decouple as before.
We now choose the variances $\sigma_{1,2}^2$ sufficiently small so that both clusters are
coherent for a given $K>1$. In particular,
\be\lbl{before}
\max\{K_c(\sigma_1^2,\alpha), K_c(\sigma_2^2,\alpha)\}< K,
\ee
i.e., the incoherent state is unstable for each cluster. Note that since the Vlasov equations are uncoupled, we can
compute the critical values $K_c(\sigma_1^2,\alpha)$ and $K_c(\sigma_1^2,\alpha)$ for each cluster separately.
Next, we keep $\sigma_2^2$ fixed and 
and increase $\sigma_1^2$ so that
\be\lbl{after}
K_c(\sigma_2^2,\alpha)< K < K_c(\sigma_1^2,\alpha).
\ee
Now the second cluster remains coherent, while the first cluster transitions to the newly stable mixing state,
thus, giving rise to a chimera state. The results of this experiment are presented in Figs.~ \ref{f.chimera_phases}{\bf a} and \ref{f.9}.

In the numerical experiments above we used the explicit expression of the stable equilibrium \eqref{equilib}
to compute the value of $\alpha^\ast$, for which the coupling coefficient $c(t)$ vanishes (cf. \eqref{find_alpha}).
Instead one can use the following adaptive scheme
to guide the system into the regime where $c(t)$ becomes very small. \footnote{Note that we are not
using the analytic equation for $\alpha^\ast$.} To this end, let us
add the following differential equation for $\alpha$:
\begin{equation}\label{eq:alpha_ode}
	\dot\alpha = \cos(U_2 - U_1 +\alpha).
\end{equation}
The right--hand side of \eqref{eq:alpha_ode} depends on the average values of $u$ computed for the
first and the second cluster
\begin{equation}
	U_1 = m^{-1}\sum_{k=1}^m u_k,\;\; U_2 = l^{-1}\sum_{k=1}^l u_k.
      \end{equation}
      Note that at any fixed point of \eqref{eq:alpha_ode}, $c(t)=\cos(U_2 - U_1 +\alpha)$ is automatically zero.
      Thus, after short transients we expect that the evolution of $\alpha$ forces $c(t)$ to become very small
      and to stay small for all future times. We verified this scenario numerically in the experiment illustrated
      in Figs.~ \ref{f.chimera_phases}{\bf b} and \ref{f.dotalpha}.


\begin{figure}
	\centering
	
	{\bf a}) \includegraphics[width = .6\textwidth]{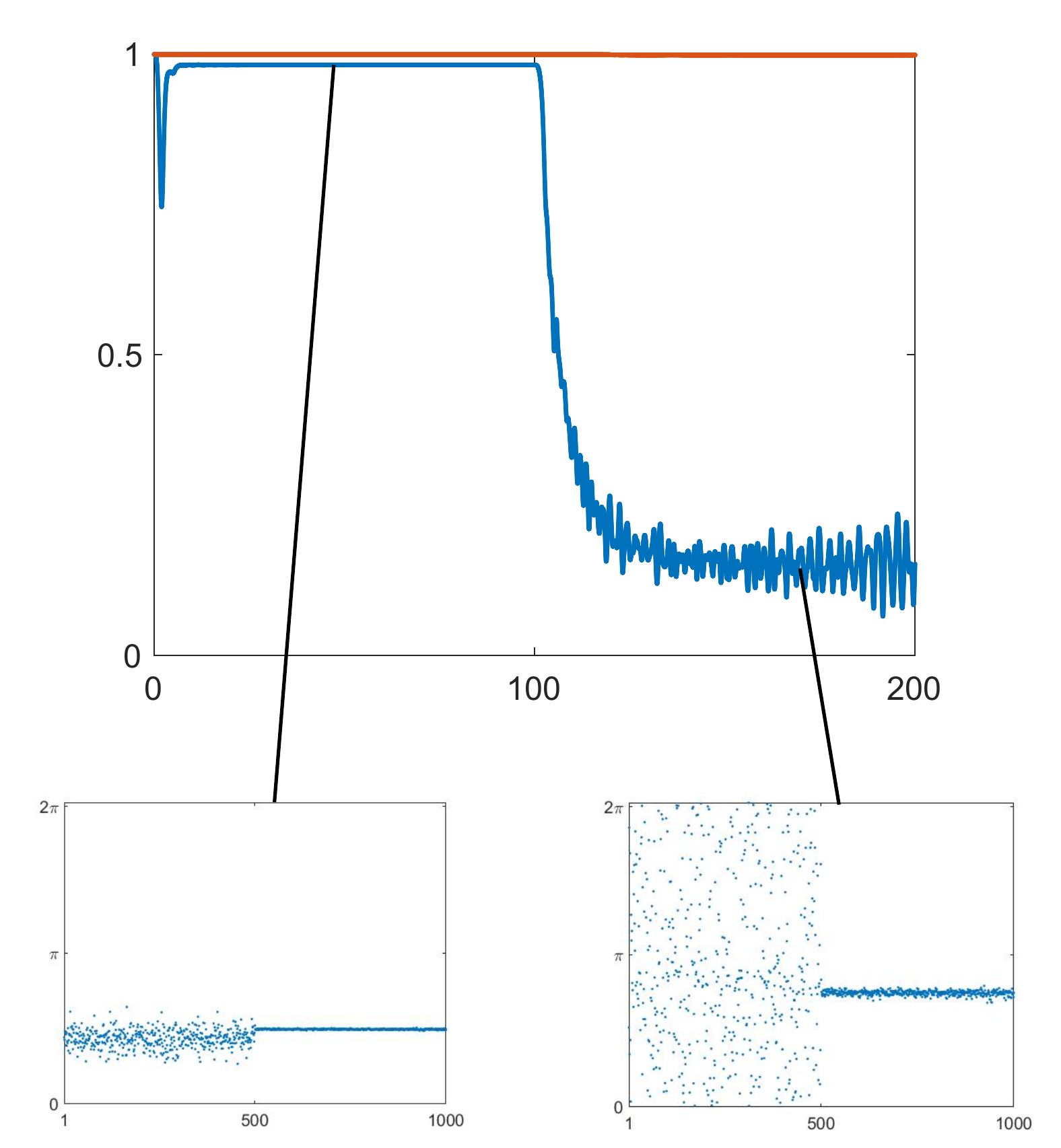}
		{\bf b}) \includegraphics[width  =.28\textwidth]{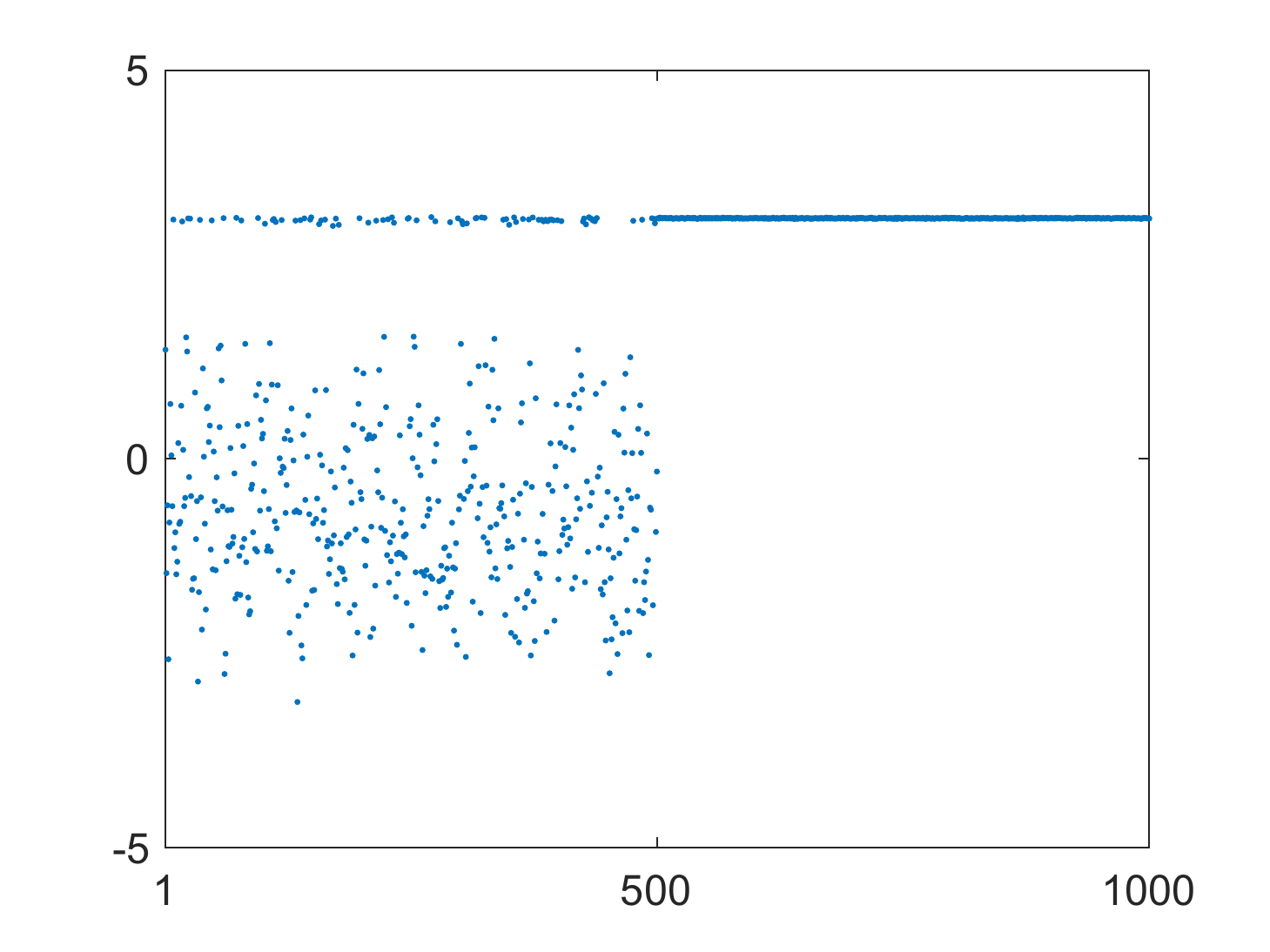}
	\caption{Starting the simulation with $\alpha=0$ fixed, we observe phase-locked solutions.
          At $t=100$ we let $\alpha$ evolve by \eqref{eq:alpha_ode} and see the emergence of a chimera state. {\bf a}) Shows the order parameter of each
          cluster over time, together with snapshots showing both phase locked and chimera states. {\bf b}) Shows a snapshot of oscillator velocities during chimera state.
          Here $\gamma=1$, $K=5$, and frequencies are chosen from $\mathcal{N}(-.5,.9)$ and $\mathcal{N}(.5,.05)$.}
        \lbl{f.dotalpha}
\end{figure}



\begin{figure}
	\centering
	{\bf a})	\includegraphics[width = .8\textwidth]{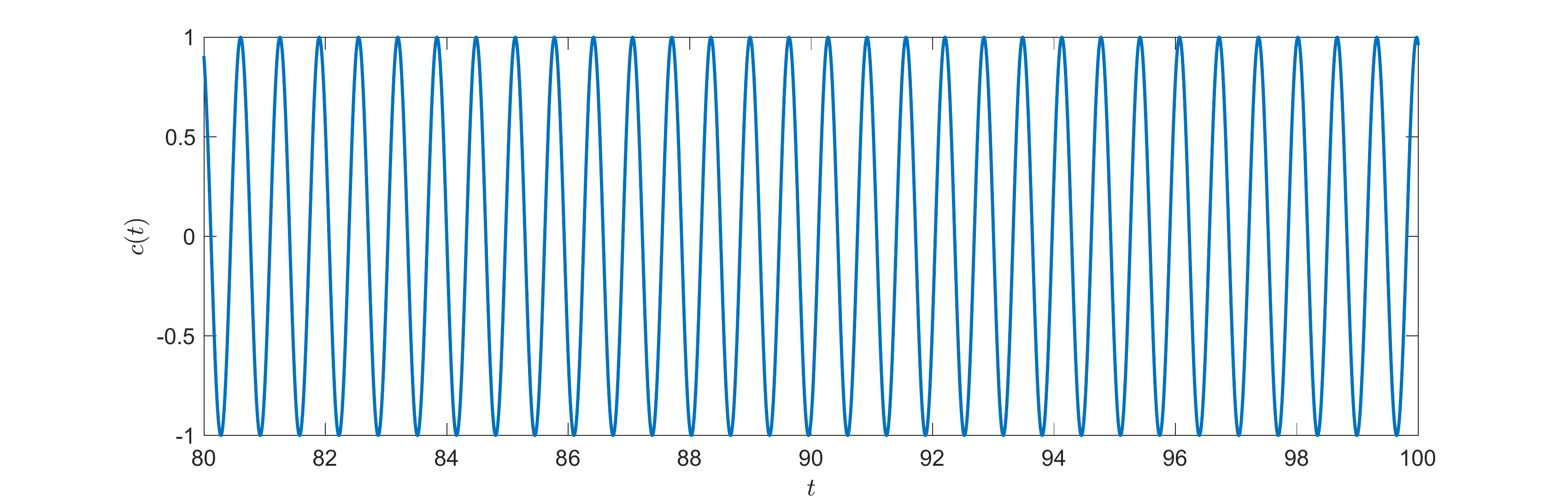}\\
		{\bf b}) \includegraphics[width = .35\textwidth]{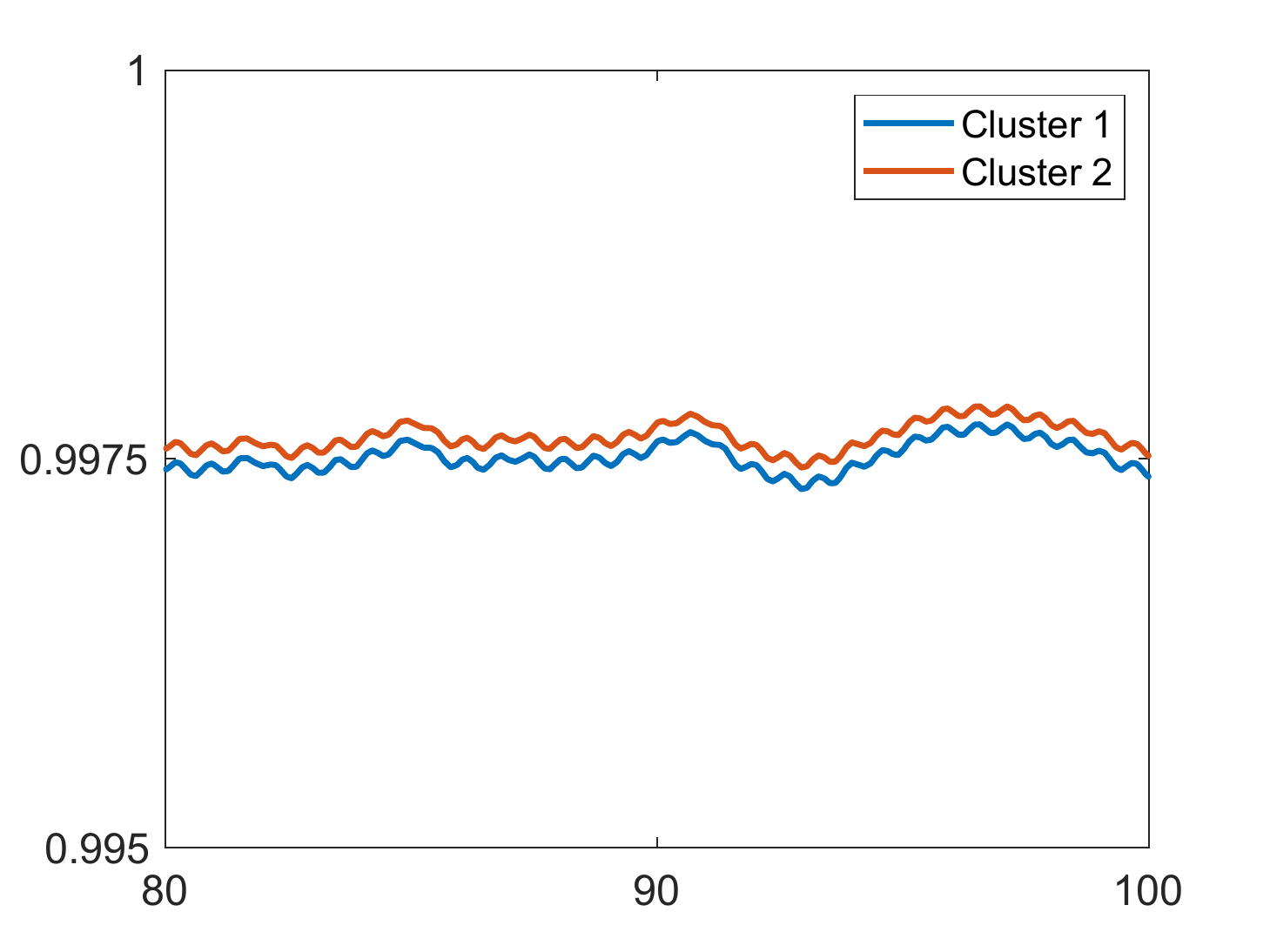}
                {\bf c}) \includegraphics[width = .35\textwidth]{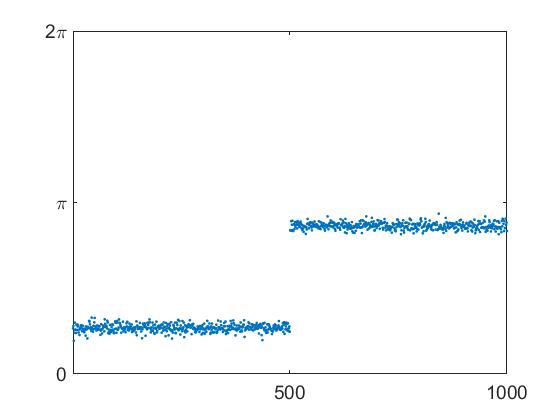}\\	
	{\bf d}) \includegraphics[width = .35\textwidth]{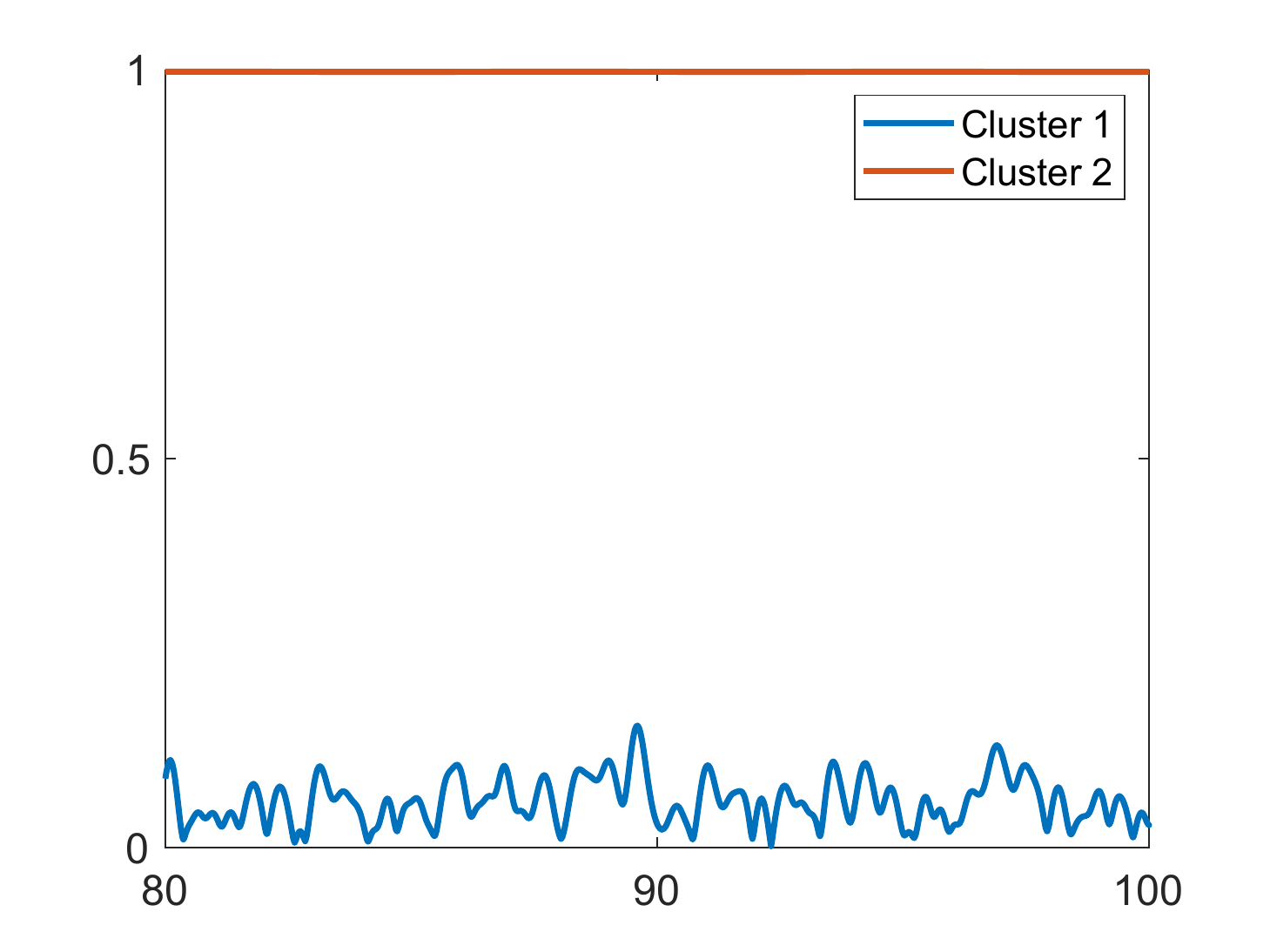}
	{\bf e}) \includegraphics[width = .35\textwidth]{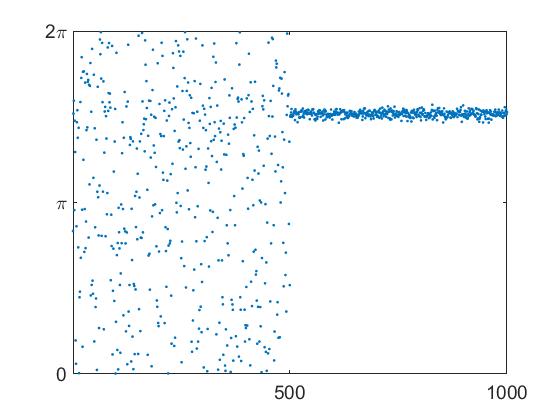}

       {\bf f}) \includegraphics[width  = .35 \textwidth]{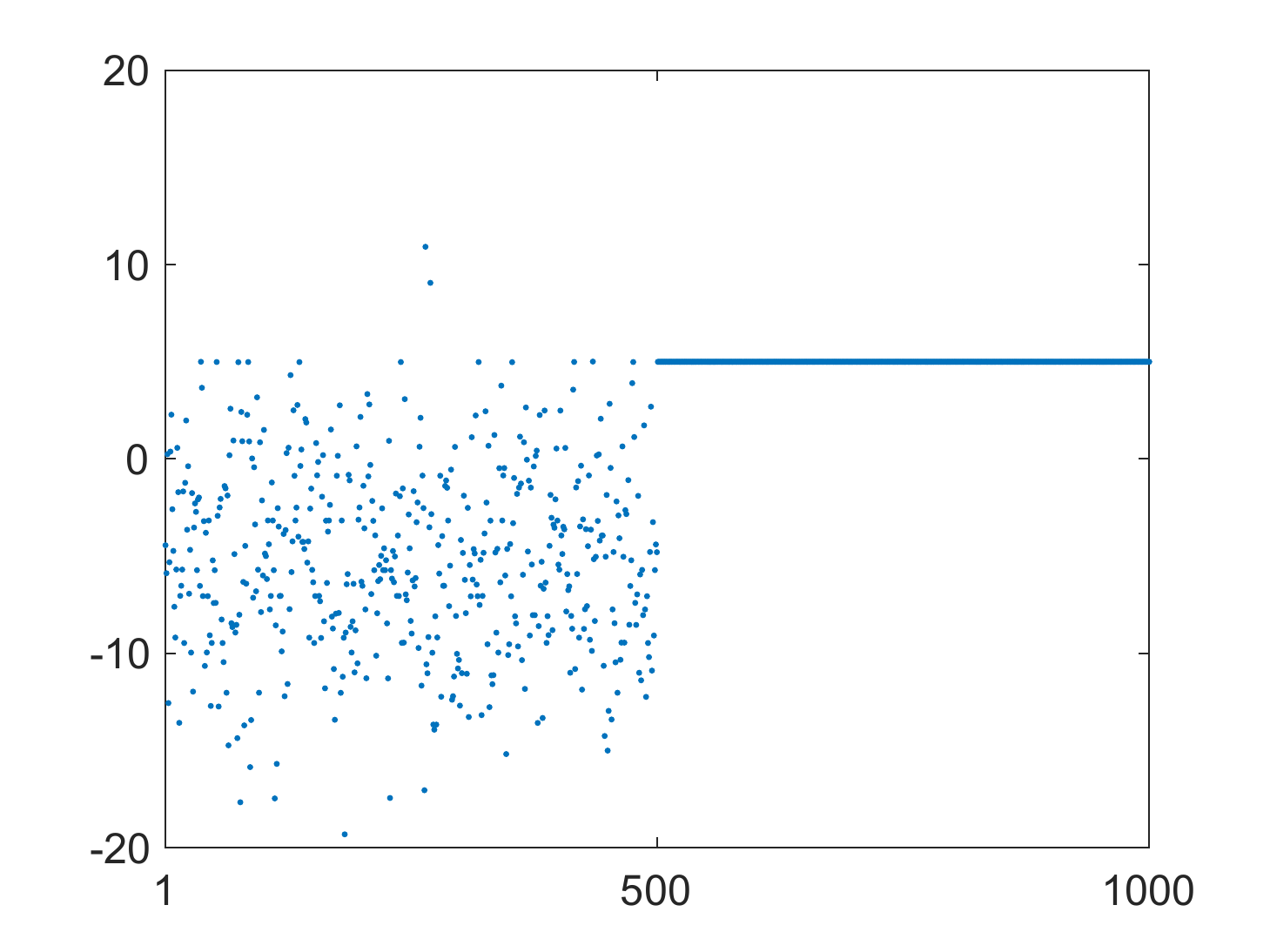}
              {\bf g}) \includegraphics[width = .3\textwidth]{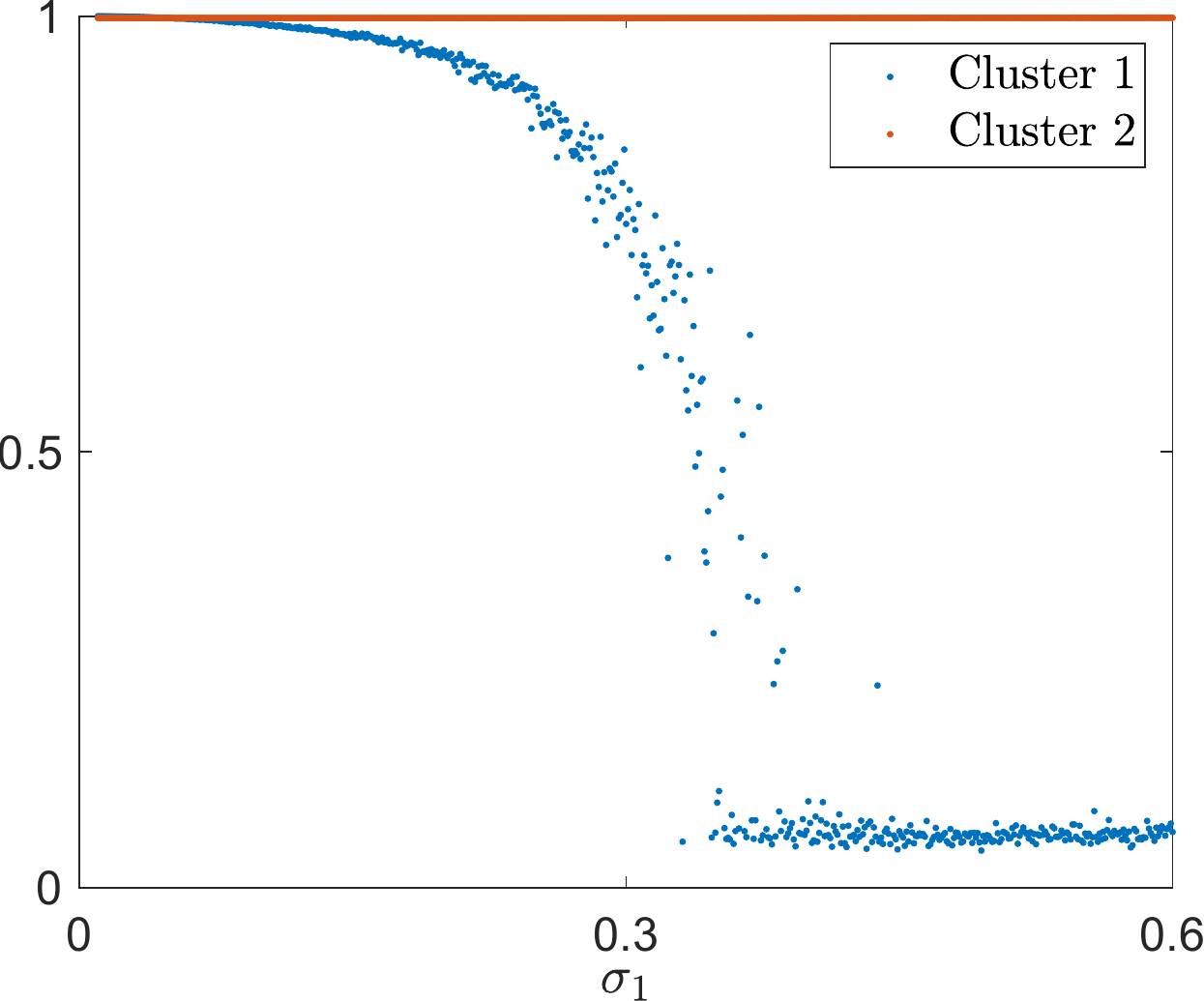}

       \caption{Taking parameters $\gamma = .1, K=1.5$ we
       	generate two coherent clusters moving in opposite directions rapidly.
       	{\bf a}) The coupling coefficient $c(t)$ oscillates rapidly around $0$.
       	Taking frequencies chosen from $\mathcal{N}(\pm .5,.05)$ we see both clusters remain synchronized.
       	{\bf b}) The order parameters for both clusters remain close to $1$ for all times, and
       	{\bf c}) a snapshot of synchronized clusters. Next taking frequencies of the
       	first cluster from $\mathcal{N}(-.5,.5)$, cluster 1 desynchronizes, while cluster 2 remains synchronous. 
       	{\bf d}) The order parameter
       	for cluster 1 rapidly converges close to zero, but the order parameter for cluster 2 remains close to
       	1 for all time. {\bf e}) A snapshot of the resultant chimera and {\bf f}) a snapshot of oscillator velocities. {\bf g}) The order parameters of each cluster when frequencies of the first cluster are taken from $\mathcal{N}(-.5,\sigma_1)$.
        }
	\label{f.11}
\end{figure}

Finally, we turn to the case when the group dynamics are driven by a limit cycle. In this case, it is easy to find the
values of parameters for which the {velocity $\dot{U}_2-\dot{U}_1$} along the limit cycle is sufficiently large and approximately constant
(see Fig.~\ref{f.pplane}{\bf a}). Then $c(t)\approx\cos(\omega t+\tau)$ for some $\omega\gg 1$ and phase shift
$\tau$ (Fig.~\ref{f.11}{\bf a}). Note that the average value of $c$ is $0$ and as before, i.e., we effectively have
uncoupled equations for
the fluctuations in the two clusters. Using this observation, we can construct numerical examples illustrating
the loss of stability of two-cluster states leading to chimera states (Fig.~\ref{f.11}).

\section{Discussion}\lbl{sec.discuss}
\setcounter{equation}{0}
The main contribution of this paper is the general framework for studying stability of clusters in the second order
KM with random intrinsic frequencies. We show that the stability of a two-cluster state depends on the stability
of the underlying group motion and the stability of coherence within each cluster. The first problem
is deterministic.
It has already been identified in the analysis of the KM with identical oscillators \cite{BBB16}.
The second problem is intrinsically
stochastic. To our knowledge, it has not been analyzed in the context of stability of clusters before.
We demonstrate that the loss of coherence in one of the clusters leads to the destabilization
of the two-cluster state. In contrast to the stability of clusters in the KM with identical oscillators
in \cite{BBB16} or the loss of stability of solitary states in \cite{JBL18}, the underlying
bifurcation is the bifurcation of the steady state of the system of Vlasov PDEs not of the
damped pendulum equation, i.e., that this is an infinite-dimensional phenomenon.
Interestingly, this leads to the creation of chimera states with one cluster staying coherent and the
other incoherent. The emerging chimera states differ from the previously reported ones in
several respects. They do not lie close to the border between the regions of the attractive and repulsive coupling
like the chimera states in the classical KM (cf.~\cite{Ome13}). They do not depend on the block
structure of the coupling (adjacency) matrix, as chimera states in \cite{Olm15, Olm15a}.
Unlike solitary states in \cite{JBL18}, they do not rely on the existence of clusters with equal velocities.
The velocities in the incoherent clusters of chimera states shown in Figs.~\ref{f.9}, \ref{f.dotalpha},
and \ref{f.11} are distributed over an interval.

The coexistence of coherence and incoherence in the homogeneous networks of coupled oscillators
has been the most intriguing feature of chimera states
since their discovery in \cite{KurBat02}. For large systems, the most comprehensive explanation
for such coexistence is based on the Ott-Antonsen Ansatz \cite{Ome13}, i.e., it applies to
a family of special solutions of the KM. The existence of the weak chimera states as defined in
\cite{AshBur15} is difficult to verify in large systems with random parameters. At the same time,
numerous modeling and experimental studies clearly demonstrate that the coexistence of coherence
and incoherence in coupled system is a universal phenomenon. In this paper, we analytically showed
the existence of two-cluster states having distinct statistical properties. The distribution
of the fluctuations in one cluster can be controlled independently from the distribution in the other
cluster. This provides a new mechanism for spatiotemporal patterns with regions with distinct statistical
properties and explains formation of chimera states shown in  Figs.~ \ref{f.chimera_phases}, \ref{f.9}, \ref{f.dotalpha},
and \ref{f.11}.

The analysis of this paper can be used to study patterns with $d>2$ clusters.
In this case, the problem of stability is reduced to a system of $d-1$ coupled pendulum equations
and $d$ coupled Vlasov PDEs. We were able to analyze certain $3-$cluster states (not presented in this
paper). However, the complexity of the problem grows rapidly with $d$. We anticipate that symmetry can
be used to understand at least certain $d$-clusters for $d>2$. Furthermore, as we remarked earlier,
our approach naturally extends to systems on more general random graphs (cf.~\cite{Med19}).
The studies of the classical KM of coupled phase oscillators made substantial contribution to our
understanding of synchronization in coupled systems \cite{Rod16}. The second order KM holds
an equal potential for the formation of clusters in large coupled dynamical systems.

\vskip 0.2cm
\noindent
{\bf Acknowledgements.}  This work was supported in part by NSF grant DMS 1715161 (to GM). Numerical simulations were completed using the high performance computing
cluster (ELSA) at the School of Science, The College of New Jersey. Funding of ELSA is provided in part by National Science Foundation OAC-1828163. MSM was additionally supported by a Support of Scholarly Activities Grant at The College of New Jersey.


\def\cprime{$'$} \def\cprime{$'$} \def\cprime{$'$}
\providecommand{\bysame}{\leavevmode\hbox to3em{\hrulefill}\thinspace}
\providecommand{\MR}{\relax\ifhmode\unskip\space\fi MR }
\providecommand{\MRhref}[2]{%
  \href{http://www.ams.org/mathscinet-getitem?mr=#1}{#2}
}
\providecommand{\href}[2]{#2}

\end{document}